\documentclass[11pt,a4paper]{article}
\pdfoutput=1
\usepackage{jinstpub}
\usepackage[english]{babel}
\usepackage{upgreek}
\usepackage{xspace}
\usepackage{graphicx}
\usepackage[capitalize]{cleveref}
\usepackage{enumitem}
\usepackage{subcaption}

\newcommand{\xmax}{\ensuremath{X_{\mathrm{max}}}\xspace}
\newcommand{\gcm}{\ensuremath{\mathrm{g/cm^2}}\xspace}
\newcommand{\sibyll}{Sibyll 2.3\xspace} 
\newcommand{\qgsII}{QGSJetII-04\xspace}

\title{Deep-Learning based Reconstruction of the Shower Maximum $\boldsymbol{X_{\mathrm{max}}}$ using the Water-Cherenkov Detectors of the Pierre Auger Observatory}

\author{\includegraphics[height=30mm]{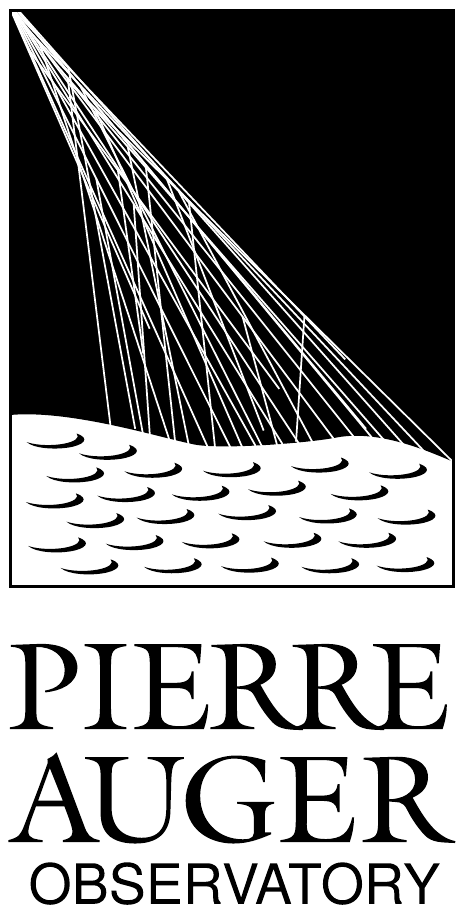}\\[3mm]The Pierre Auger Collaboration}
\affiliation{Av.\ San Mart\'{\i}n Norte 306, 5613 Malarg\"ue, Mendoza, Argentina}

\emailAdd{auger\_spokespersons@fnal.gov}

\abstract{
The atmospheric depth of the air shower maximum \xmax is an observable commonly used for the determination of the nuclear mass composition of ultra-high energy cosmic rays.
Direct measurements of \xmax are performed using observations of the longitudinal shower development with fluorescence telescopes.
At the same time, several methods have been proposed for an indirect estimation of \xmax from the characteristics of the shower particles registered with surface detector arrays.
In this paper, we present a deep neural network (DNN) for the estimation of \xmax. The reconstruction relies on the signals induced by shower particles in the ground based water-Cherenkov detectors of the Pierre Auger Observatory.
The network architecture features recurrent long short-term memory layers to process the temporal structure of signals and hexagonal convolutions to exploit the symmetry of the surface detector array.
We evaluate the performance of the network using air showers simulated with three different hadronic interaction models. Thereafter, we account for long-term detector effects and calibrate the reconstructed \xmax using fluorescence measurements. Finally, we show that the event-by-event resolution in the reconstruction of the shower maximum improves with increasing shower energy and reaches less than $25$~\gcm at energies above $2\times 10^{19}$~eV.
}

\arxivnumber{2101.02946}

\dedicated{\flushleft\textnormal{\textsc{%
Published in JINST as \href{https://doi.org/10.1088/1748-0221/16/07/P07019}{DOI:10.1088/1748-0221/16/07/P07019}
\\
Report number: FERMILAB-PUB-21-084-AD-AE-SCD-TD
}}}
\begin{document}

\maketitle
\flushbottom

\section{Introduction}
The depth of the air shower maximum \xmax induced by ultra-high energy cosmic rays is a key observable to estimate the mass of primary cosmic particles (see e.g. \cite{kampert:2012mx} for a review).
In this work we introduce a new method to reconstruct \xmax from signals recorded by water-Cherenkov detectors (WCDs), whose duty cycle is almost $100\%$, unlike fluorescence telescopes, which can only operate during moonless nights with a duty cycle of about $15\%$.

\begin{figure*}[t!]
    \begin{centering}
        \begin{subfigure}[b]{0.49\textwidth}
            \begin{centering}
                \includegraphics[height=5.75cm]{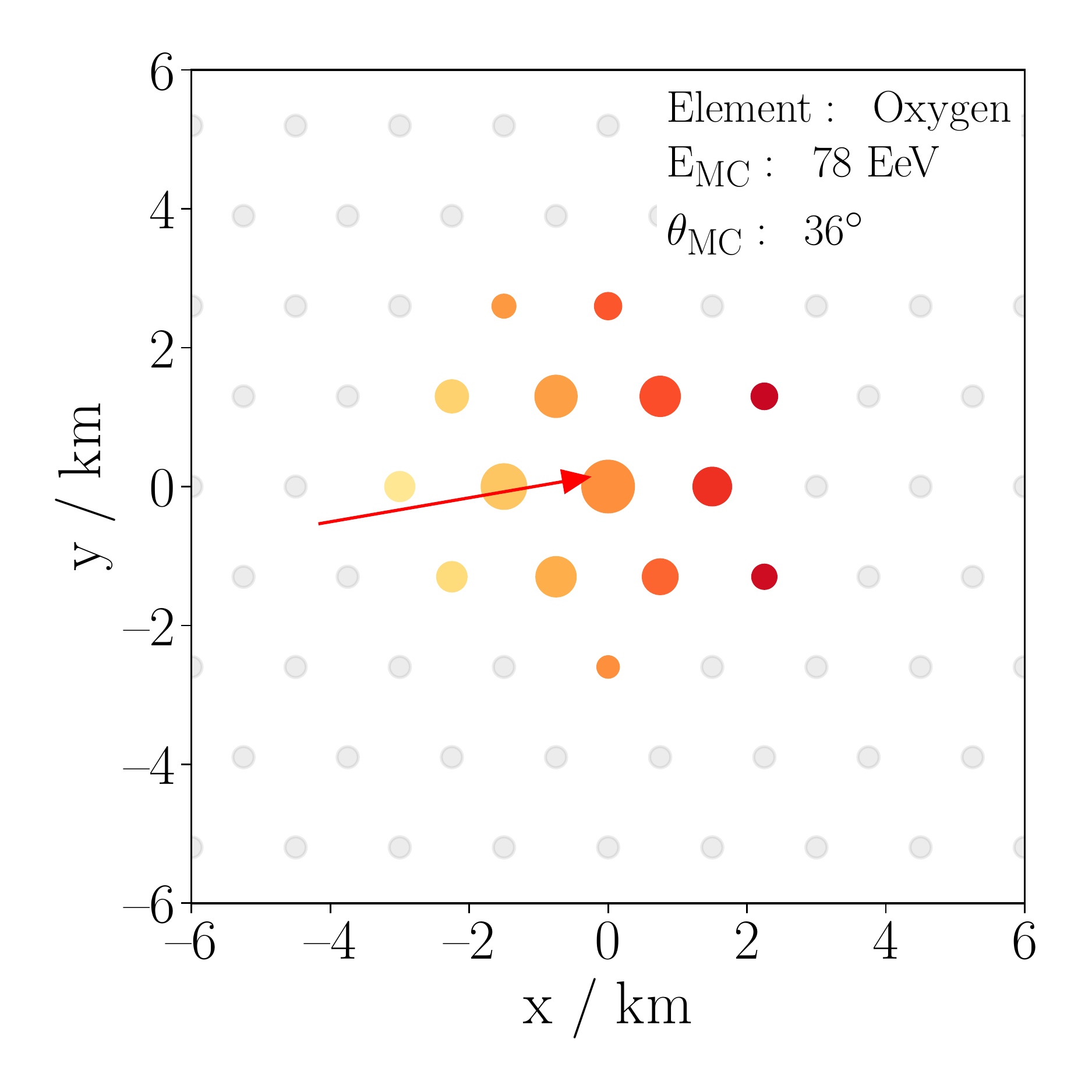}
                \subcaption{}
                \label{fig:footprint}
            \end{centering}
        \end{subfigure}
        \begin{subfigure}[b]{0.49\textwidth}
            \begin{centering}
                \includegraphics[height=5.75cm]{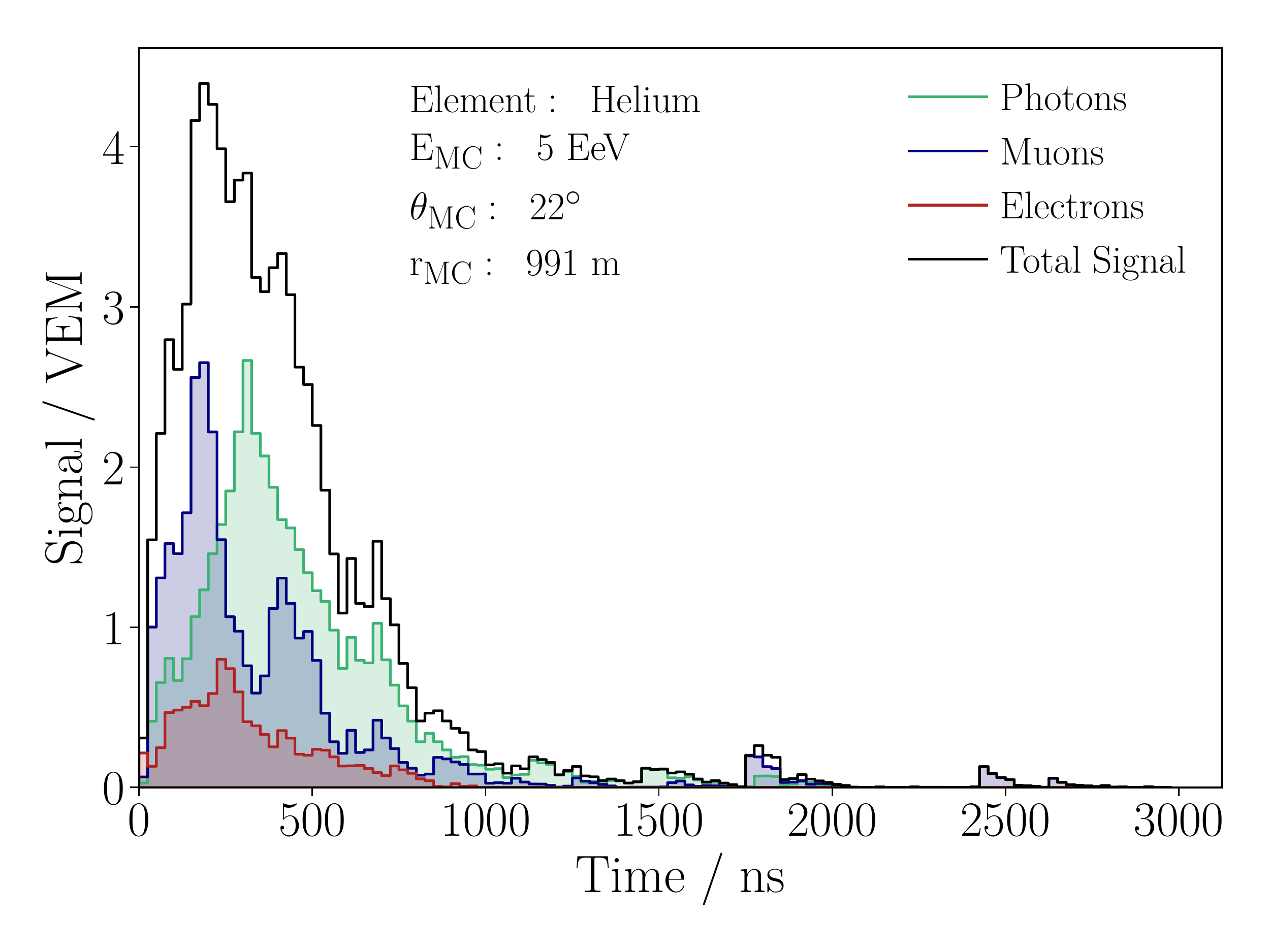}
                \subcaption{}
                \label{fig:trace}
            \end{centering}
        \end{subfigure}
        \caption{(a) Simulated signal pattern measured by the surface detector. The marker sizes indicate the amount of measured signal and the colors represent the arrival time of the shower at a given station (yellow=early, red=late). The arrow denotes the projection of the shower axis on the surface and its tip the shower core. (b) Simulated signal trace of a cosmic-ray event measured at a surface-detector station at a distance of about $1000$~m to the shower core. Different colors indicate signals from different shower components.}
        \label{fig:data}
    \end{centering}
\end{figure*}

The Pierre Auger Observatory~\cite{auger_nim} was constructed to study cosmic rays at the highest energies with high accuracy. With its instrumented area of $3000$~km$^2$ the observatory is sufficiently large to measure the rare high-energy events up to $10^{20}$~eV with adequate statistics.
Designed as a hybrid instrument, on the one hand the development of the air showers is directly followed by observing the fluorescence light emitted by nitrogen molecules in the air with $27$ telescopes, constituting the fluorescence detector (FD).
As measurements using the fluorescence technique are only possible during moonless nights it is used to collect data for an in-depth understanding of the air showers~\cite{fd_xmax, fd_xmaxI, fd_xmaxII, fd_xmaxIII} with reduced statistics. On the other hand, secondary particles of the air showers are detected by $1600$ water-Cherenkov Detectors (WCDs) located on the ground, which form the surface detector (SD). The detectors are located at a distance of $1.5$~km from each other in a hexagonal configuration (Fig.~\ref{fig:footprint}) and measure the time-dependent density of secondary shower particles.
The measurement is based on Cherenkov light emitted by the relativistic particles as they pass through each water-filled detector. This light is recorded by three photomultiplier tubes (PMTs) attached to the top of the tank and processed using flash analog-to-digital converters with a sampling rate of $40$~MHz, translating to bins with a width of $25$~ns.

For the determination of the air shower direction and energy with the WCDs there are already established precise reconstruction methods~\cite{sd_reco} that use the arrival times of the shower particles and the spatial extent of the air shower. Reconstructing the depth of maximum of the shower, on the other hand, is a major challenge. In contrast to the fluorescence telescopes, there is no direct observation of the shower development in the atmosphere. Instead, the shower properties are encoded in the signal traces induced by the secondary particles traversing the WCDs.

During the longitudinal development of the air shower, the relative abundances of hadrons, muons, electrons and photons change in addition to geometric dependencies such as the distance to the shower core or the zenith angle. The signal traces generated by these individual components in the three PMTs of the WCDs are different and thus provide the possibility to extract information about the depth of the shower maximum \xmax. Whereas muons do not interact with the atmosphere and reach the detector usually in earlier times, the electromagnetic component is strongly shielded and reach the detector later and more spread in time. See Fig.~\ref{fig:trace} for a simulated example trace with muons and electromagnetic shower particles.

Experimental proof of the successful use of the signal characteristics to determine the depth of maximum of the shower has already been achieved.
It was demonstrated~\cite{universality-1, universality-2} that using shower universality~\cite{shower_univ, univ_lipari} $X_{\mathrm{max}}$ can be measured with good accuracy by decomposing the signal traces using templates of typical waveforms induced by the different shower components using the universality of signals with respect to the shower development stage.
Furthermore, the correlation between the risetimes of the signals in the WCDs was used to determine \xmax~\cite{Aab:2017cgk}.
With this method the average composition of UHECRs was determined up to $100$~EeV. The measurement up to such high energies has been only made possible using the large exposure and high reconstruction efficiency of the SD, increasing the statistics by a factor 25 above $3$~EeV (factor 12 above $30$~EeV) when compared to the FD~\cite{delta_icrc}.

Our new method aims to measure, beside the average composition, the mixing of the composition $\sigma(\xmax)$ with high statistics by exploiting mass-sensitive information on an event-by-event basis. This opens up possibilities for event-based anisotropy studies using information on the cosmic-ray mass.

This method to reconstruct $X_{\mathrm{max}}$ using the measured signals of the WCDs is based on a deep neural network (DNN), whose architecture and optimization was developed specifically for the measurement conditions at the Pierre Auger Observatory. The network is a further development of \cite{aixnet} and takes into account current developments in machine learning, especially elements from speech recognition and utilization of spatial symmetries.
\newpage
In the hybrid network architecture, all time intervals of the signal traces of the PMTs are first scanned using a recurrent subnetwork. Thereby each signal trace is characterized by a total of $10$ ``machine-learned features``. In the following network layers these observables are merged with the particle arrival times at the respective WCD locations. When processing the signals at the given positions, the hexagonal symmetry of the observatory is generically mapped into the network architecture.
The network parameters for forming the $10$ observables to characterize the signal traces and for combining all available information are adjusted in a training procedure. The most important tools of this optimization are the training data and the objective function, which is minimized during network training.

This work is structured as follows. First, we specify the data sets for both the simulation studies and measured Auger hybrid data, which include information from the FD for validation purposes. We explain in detail how the simulated data are prepared and augmented for the optimization of the network parameters and the reconstruction of \xmax. After that, we describe in detail the architecture and training of the deep network. Then we show the \xmax reconstruction performance of the network on simulated data as a function of energy, zenith angle, mass of the primary particle, and the effect of using two hadronic interaction models different from the one used in the training. Finally, we verify the capabilities of the network by direct comparison of the measured maximum shower depth \xmax of the network and of the FD. We correct for detector-aging effects resulting from long-term operation of the observatory. Subsequently, we calibrate the absolute \xmax value of the network output, and determine the \xmax resolution of the network as a function of the primary energy.

\section{Data sets and their preparation}

The measured air shower footprint consists of a characteristic pattern of several triggered WCDs arranged in a hexagonal grid (see Fig.~\ref{fig:footprint}). Using three PMTs each triggered station measures the time-dependent density of particles encoded in three signal traces. An example of a simulated signal trace is shown in Fig.~\ref{fig:trace}.

The basic idea is to provide the network as input the raw data of a measured cosmic-ray event. The raw information for each triggered station consists of three signal time traces, the station position and the time of the first shower particles arriving at the station.

For successful adjustment of the network parameters, careful preparations of the data sets from simulation campaigns used for the optimization process are crucial. For example, the parameters can be set more easily if the numerical values of the input variables do not vary considerably. Therefore, both the amplitudes and the time values of the WCD-signal traces are re-scaled prior to their input into the network. Also, typical conditions when operating the observatory need to be included in the simulated training data used for adjusting the parameters.

Below we specify the data sets from simulation campaigns, and data with information from the FD used to validate the \xmax reconstruction of the deep neural network.

\subsection{Simulation libraries} 

For generating training and validation data we use CORSIKA 7.6400~\cite{corsika} for the simulation of extensive air showers. The Pierre Auger Offline software~\cite{offline} is used for the simulation of the detector responses, which is based on GEANT4~\cite{geant4}, and the reconstruction of the air shower parameters~\cite{auger_nim}. The CORSIKA showers are reused $5-30$ times at other locations of the observatory along with a re-simulation of the detector responses. The simulation library consists of events for primary hydrogen, helium, oxygen and iron, uniformly distributed in azimuth angle and following a zenith angle distribution flat in $\cos^2\theta$ with $0^\circ\leq \theta < 65^\circ$. 
The energy range covers $1-160$~EeV (1 EeV = $10^{18}$~eV) and follows a spectrum of $E^{-1}$. For the training of the network we use exclusively simulations with the EPOS-LHC hadronic interaction model~\cite{epos-lhc}. Overall, the training set contains around $400{,}000$ events.

For testing the performance we prepare a test set with EPOS-LHC showers containing $50{,}000$ events and two sets of showers simulated using the hadronic interaction model \qgsII~\cite{qgsjetII} and \sibyll~\cite{sibyll2.3c} containing $450{,}000$ events each. In contrast to the training data for the test sets we limit the energy range to $3-160$~EeV and the zenith angle range to $0-60^{\circ}$, where the standard SD reconstruction shows no significant reconstruction biases~\cite{auger_nim}. The extended phase space during training prevents incorrectly reconstructed energies and zenith angles at the edges of the phase space.

\subsection{Hybrid dataset\label{sec:hybrid_data}}

The accuracy of the deep-learning based reconstruction of \xmax is examined using a high-quality set of hybrid data where nearly unbiased \xmax measurements are performed using the fluorescence technique. A complete description of the hybrid reconstruction and the high-quality event selection can be found in~\cite{fd_xmax}.

Here, events are only kept if \xmax falls into the FD field of view and the fraction of the shower profile inside the field of view is large enough to allow an unbiased measurement of \xmax.
Additionally, we discard events below the full trigger efficiency of the surface detector at roughly $3.16$~EeV to prevent biased reconstructions. Further, we remove events (less than $<0.5\%$) with area over peak (A/P) values~\cite{longterm}, given in units of $25$~ns, which are very low $(<2.7)$ or high $(> 3.5)$ and cannot be corrected using a linear function (see section \ref{sec:aop}). To guarantee a precise SD reconstruction, we reject events with broken or non-existent stations in the first hexagon formed around the station with the largest signal~\cite{trigger}. After the selection $3109$ events, collected from 1 January 2004 to 31 December 2017, remain for the evaluation of the DNN performance.

\subsection{Pre-processing of data} 
\label{prepro}
Footprints of air showers on the ground can reach sizes of several tens of square kilometers. To reduce the memory consumption and increase generalization capacities of the DNN, we use only the information from a fixed size of the SD consisting of a sub-array of $13 \times 13$ WCDs around the WCD with the highest signal.

For air showers with zenith angles below $\theta=60^{\circ}$ more than $99 \%$ of the triggered stations are contained within this sub-array. We expect the effect of enlarging the sub-grid to be negligible as only stations with very low signals lie outside the sub-grid. 

As the positions of the WCDs feature a hexagonal grid several representations of the local neighborhood are possible~\cite{aixnet}. Our algorithm is based on the so-called axial representation.

\paragraph{Signal traces}
The total number of particles measured in each detector is expected to approximate a power law as a function of the distance to the shower axis. Hence, we use a logarithmic re-scaling of the signal traces $S_i(t)$:

\begin{equation}
\tilde{S}_i(t) = \frac{\log_{10}[S_i(t) / \mathrm{VEM} + 1]}{\log_{10}[S_\mathrm{norm} / \mathrm{VEM}+1]}
\end{equation}
Here VEM (vertical equivalent muon) is a unit equal to the signal
induced by a muon traversing the WCD vertically and $S_\mathrm{norm}=100$~VEM normalizes signals of $100$~VEM to unity. To allow for positive values only and to leave contributions with $S=0$ unchanged we use $S(t) + 1$.

\paragraph{Particle arrival times}
The curvature of the shower front is reflected in the arrival times $t_{0,i}$ of the first shower particles at the SD stations.
For each triggered station in an event, the particle arrival time is normalized with respect to the arrival time $\tau_{\rm center}$ measured at the station with the largest signal and the standard deviation $\sigma_{t, {\rm train\;data}}$ of the arrival times of the complete training set:

\begin{equation}
\label{eq:arr_time}
\tilde{t}_{0,i} = \frac{t_{0,i} - \tau_{\rm center}}{\sigma_{t, {\rm train\;data}}}
\end{equation}

\paragraph{Station states}
Air showers falling close to the edges of the detector grid can have footprints with a substantial fraction lying outside the surface-detector grid. Also, in rare cases, a WCD within the $13 \times 13$ sub-grid may not provide a signal owing to technical problems. To provide the information in the network training if the absence of a signal originates from air shower physics or from detector effects, we add a feature map for the given sub-grid encoding the station states as additional input ($1=$ ready to measure, $0=$ broken or missing in the grid).

\paragraph{Network prediction}
During the training of the network the prediction (output) of the neural network is compared to the $true$ labels of the dataset in each iteration. As the input values of the network lie approximately in the range of $[-1;1]$ and the weights of the model are properly initialized, the expected output values of the network follow approximately a Gaussian with standard deviation $\sigma = 1$ and mean $\mu = 0$. 

To speed-up the learning process we also normalize the label values to lie in the same range using standard normalization ($\sigma = 1, \mu = 0$). Instead of normalizing the label values directly, the normalization is implemented as the final layer in our network for the \xmax output allowing us to monitor the learning process in physics quantities.

\subsection{Augmentation of simulated data} 
\label{augmentation}
Depending on the sub-grid, the detector states of the WCDs vary on an event-by-event basis. Furthermore, there are differences between detector states in data and simulation.
Even small deviations between simulation and data can have a non-negligible influence on the network prediction if there are differences in the phase space. For example, a broken WCD would create a hole in the measured footprint which then distracts the reconstruction if the model is not trained on such detector conditions. To make the algorithm robust to distortions and fluctuations, we implemented an on-the-fly augmentation. Hence during the training every particular simulated event will feature in each repeated iteration different detector states. Note that this augmentation is exclusively used in the network training procedure and is not applied when evaluating the network.

\paragraph{Photomultiplier tubes without response}
During detector operation, one or more of the three photomultiplier tubes (PMTs) of a WCD may not respond. To make the network robust to missing signal traces, we randomly mask signal traces during the training. The number of broken PMTs is approximated as Poisson distribution and was tuned to follow the data distribution.

\paragraph{Water-Cherenkov detectors without response}
As already stated above, a map of detector stations without response is used describing the current station states. During the network training, detector stations are randomly marked as broken, by setting all measured properties to zero, which is additionally accounted for in the map of stations without response. The number of broken stations is approximated as Poisson distribution and was tuned to follow the distribution of measured data.

\paragraph{Saturated signal traces}
Shower geometries with  a shower core close to a detector station can exceed the maximum signal that can be quantitatively recorded. This effect is even more frequent at the highest energies, where $50\%$ events of air showers with $50$~EeV exhibit typically one saturated detector station.
Each surface-detector station is calibrated for the value of $1$ VEM in units of the hardware every $60$~s. This value changes for each detector station and PMT owing to individual hardware properties and other external influences. Since the number of units is limited, the saturation value is time dependent and differs for each detector station~\cite{sd_reco}.

To make the network robust to different saturation values, we sample the maximum measurable values from a Gaussian distribution adapted by typical saturation values observed in data. In detail we make use of an one-sided truncated Gaussian to prevent very small and negative saturation values. Independently for each PMT, we finally clip the unsaturated trace provided by the simulation according to the sampled saturation value.

\section{Deep neural network for reconstructing the shower maximum}

Two challenges are to be mastered for reconstructing \xmax from the signal traces of the water-Cherenkov Detectors:
\begin{enumerate}
    \item To develop a network architecture that is optimally suited for the situation of the Pierre Auger Observatory and exploits all symmetries in the recorded data,
    \item To adapt about $10^6$ network parameters in a way that the correct shower depth is reconstructed under realistic operation conditions of the observatory.
\end{enumerate}
In the following both the network architecture and the network training are described in detail.

\subsection{Architecture}

\begin{figure}[b!]
    \begin{center}
        \includegraphics[width=0.62\textwidth]{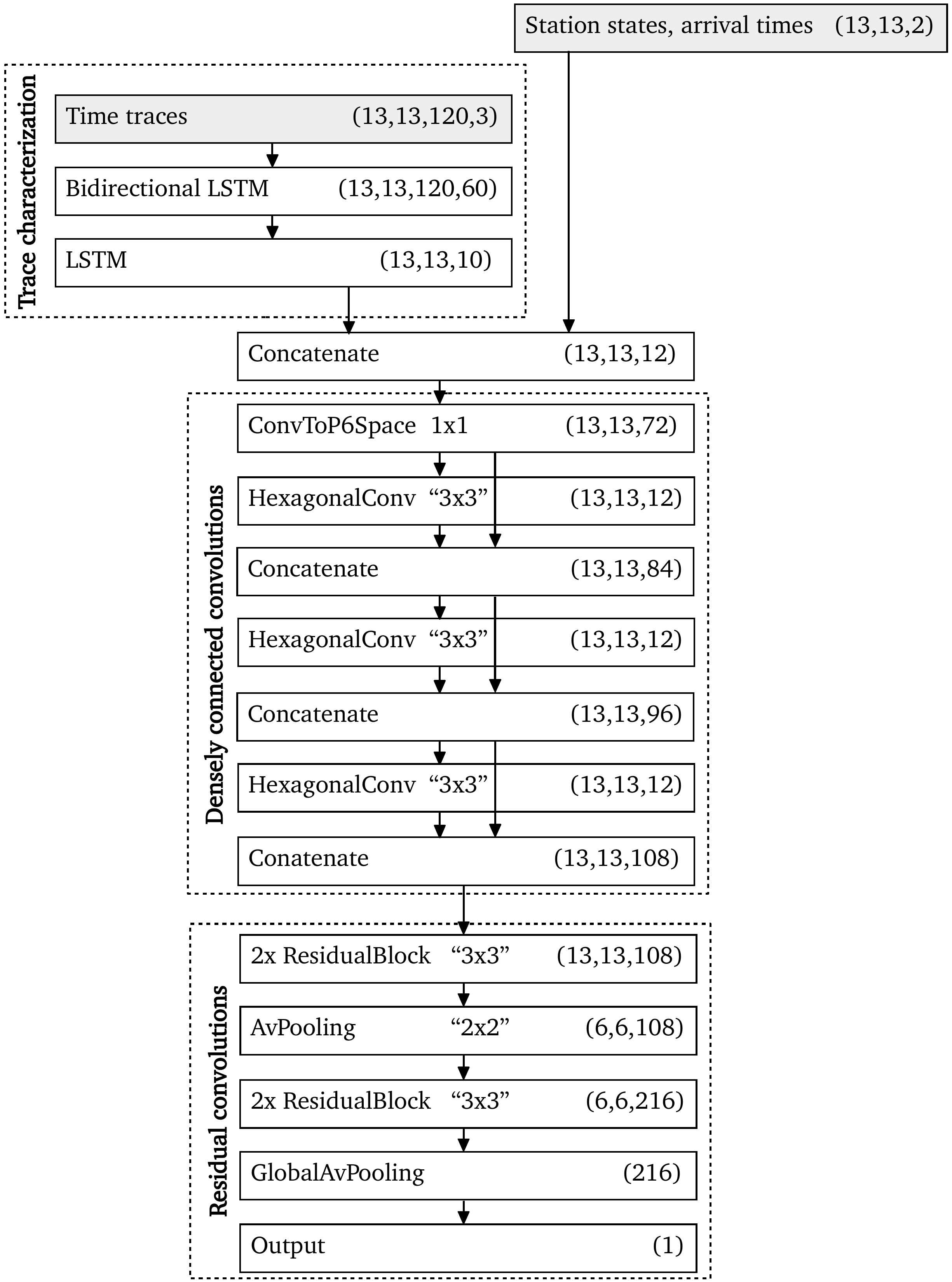}
        \caption{Architecture of the deep neural network used to reconstruct \xmax. Convolutional operations are shown with their kernel sizes, e.g. a "$3\times 3$" hexagonal convolution translates to a kernel covering one hexagon (7 WCD locations). The numbers in brackets denote the output shapes of the operations.}
        \label{fig:architecture}
    \end{center}
\end{figure}

To reconstruct the shower maximum \xmax using a deep neural network we designed an architecture featuring hexagonal convolutions~\cite{convolution, hexaconv} and recurrent layers illustrated in Fig.~\ref{fig:architecture}. We observed that using an architecture which supports the symmetry of the hexagonal grid of the detectors not only boosts the performance, but further makes the network robust to small fluctuations in the data which results in an improved generalization performance.

The input for the neural network is a tensor containing the signal time traces, the map of the arrival times and a map of the station states. For connecting the different physics quantities and extracting the maximum information from the data, we adapted state-of-the art methods of computer vision and speech recognition.

We introduce a prior on physics relationships, hence learning first to extract information of the shower development using the signal traces before forming features of space and time. The architecture can be split into three parts, discussed in the following.

\paragraph{Detector-wise recurrent processing of signal traces}
In the first part of the architecture, the signal trace is processed. To resolve long- and short-term correlations in the signal trace we make use of $1$ bi-directional layer and $1$ one-directional layer of LSTM cells~\cite{lstm}. Using bidirectional layers allows to connect not only the past time steps with the current step but additionally all future steps.

The most important concept is the station-wise sharing of the recurrent layers. As the physics of the measured traces is the same for each station, we share the trace network over all stations in the $13 \times 13$ sub-grid. Hence, exactly the same parameters are used in each station.

Note that this concept still allows the network to extract many very different features and not only a distinct characteristic of the trace. Although the same feature is extracted in each station (e.g. slope of the rising edge) each respective feature value will vary (e.g. slope differs in each station) within one event.

In this sub-network, the tensor of signal traces with a length of $120$ time steps is reduced to $10$ feature maps encoding the longitudinal shower development.
Note that these $10$ features, which are extracted for each station, are not hand-designed but are learned by the network to be particularly useful for the reconstruction of \xmax. As every station uses the same network, the resulting $10$ feature maps share the same physics meaning, allowing to use convolutions in the spatial dimensions in the following part.

\paragraph{Densely-connected hexagonal convolutions}
After analyzing the signal trace, the additional information, consisting of the arrival times of the first shower particles at the WCDs and of detector stations without response, is concatenated with the tensor already processed.

In this second part of the architecture, features relating space and time are extracted using convolutional operations on the WCD positions. Before applying hexagonal convolutions the representation needs to be transformed into the space of hexagonal rotations (P6), which is done individually for each feature map. In this space each filter holds $6$ feature maps, one for each orientation~\cite{hexaconv}.

The main concept of this block are densely-connected convolutions~\cite{densenet}, which provide already extracted features in each subsequent layer. The resulting connections stabilize the training process by feature-reuse and an improved propagation of gradients.
The block consists of $3$ layers with ELUs~\cite{elu} (exponential linear units) as activations and with kernels of size $7$ covering one hexagon.

\paragraph{Residual reconstruction modules} 
The last part of the network features residual modules introduced by ResNet~\cite{resnet}. In the standard layout of the AixNet architecture~\cite{icrc19_aixnet} for each reconstruction task (shower axis, shower core, energy and \xmax) an individual subpart exists. Each part consists of $2$ residual modules with $2$ layers each, ReLUs (rectified linear units) as activation functions and pooling operations between the blocks. The final layer of the \xmax block predicts the shower maximum after rescaling back to physics quantities as described in section~\ref{prepro}.

\subsection{Training}
For training of the network we use $400{,}000$ simulated air shower events with a mixed composition of hydrogen, helium, oxygen and iron (in equal parts). During the training we use the data augmentation as described in section~\ref{augmentation} to increase the amount of data and to improve the generalization capacities of the network. 

To penalize composition biases we use an element-wise mean squared error (MSE) as objective function:

\begin{align}
    \mathcal{L} & =  \sum_{Z = \mathrm{H, He, O, Fe}} \langle \operatorname{MSE}(\hat{X}_{\mathrm{max},Z}) \rangle =  \sum_{Z = \mathrm{H, He, O, Fe}} \langle (\hat{X}_{\mathrm{max},Z}-X_{\mathrm{max},Z})^2 \rangle \\
    & = \sum_{Z = \mathrm{H, He, O, Fe}} \operatorname{Var}(\hat{X}_{\mathrm{max},Z}-X_{\mathrm{max},Z}) +  \langle \hat{X}_{\mathrm{max},Z} - X_{\mathrm{max},Z} \rangle^2 .
\end{align}
Here $\hat{X}_{\mathrm{max},Z}$ is the predicted shower maximum and $X_{\mathrm{max},Z}$ the label, for the respective element $Z$. $ \operatorname{Var} (\;\cdot\;)$ represents the batch-wise variance and $ \langle \;\cdot\; \rangle $ the batch-wise expectation value. The network and the training is implemented using Keras~\cite{keras} and the TensorFlow~\cite{tensorflow} backend.

We train our network using an Nvidia Geforce 1080 GTX GPU for $150$ epochs with a batch size of $48$, which takes about $60$~h. We further stop the training if the objective function applied to validation data does not decrease after $10$ epochs. We use the ADAM optimizer~\cite{adam} with an initial learning rate of $0.001$ and a decay of $2.5 \cdot 10^{-6}$. Furthermore, we reduce the learning rate by multiplying by a factor of $0.7$ if the objective function evaluated on the validation data did not decrease in the recent $4$ epochs.

As part of the work, a wide variety of network architectures was tested. These tests included various changes of each of the three network parts. In particular, different types of recurrent layers and various convolutional approaches were tested. The final model which utilizes LSTMs for the processing of signals and hexagonal convolutions for analyzing the spatial relations showed the best performance and stability. In addition, we found that this design is relatively robust to small changes in the architecture. A reasonable change of the hyperparameters, e.g. the number of layers or the number of features extracted by the trace network, had no significant impact on the final result.

\newpage

\begin{figure*}[tb!]
    \begin{centering}
        \begin{subfigure}[b]{0.495\textwidth}
            \begin{centering}
                \includegraphics[width=0.99\textwidth, trim={0 0.45cm 0 0}, clip]{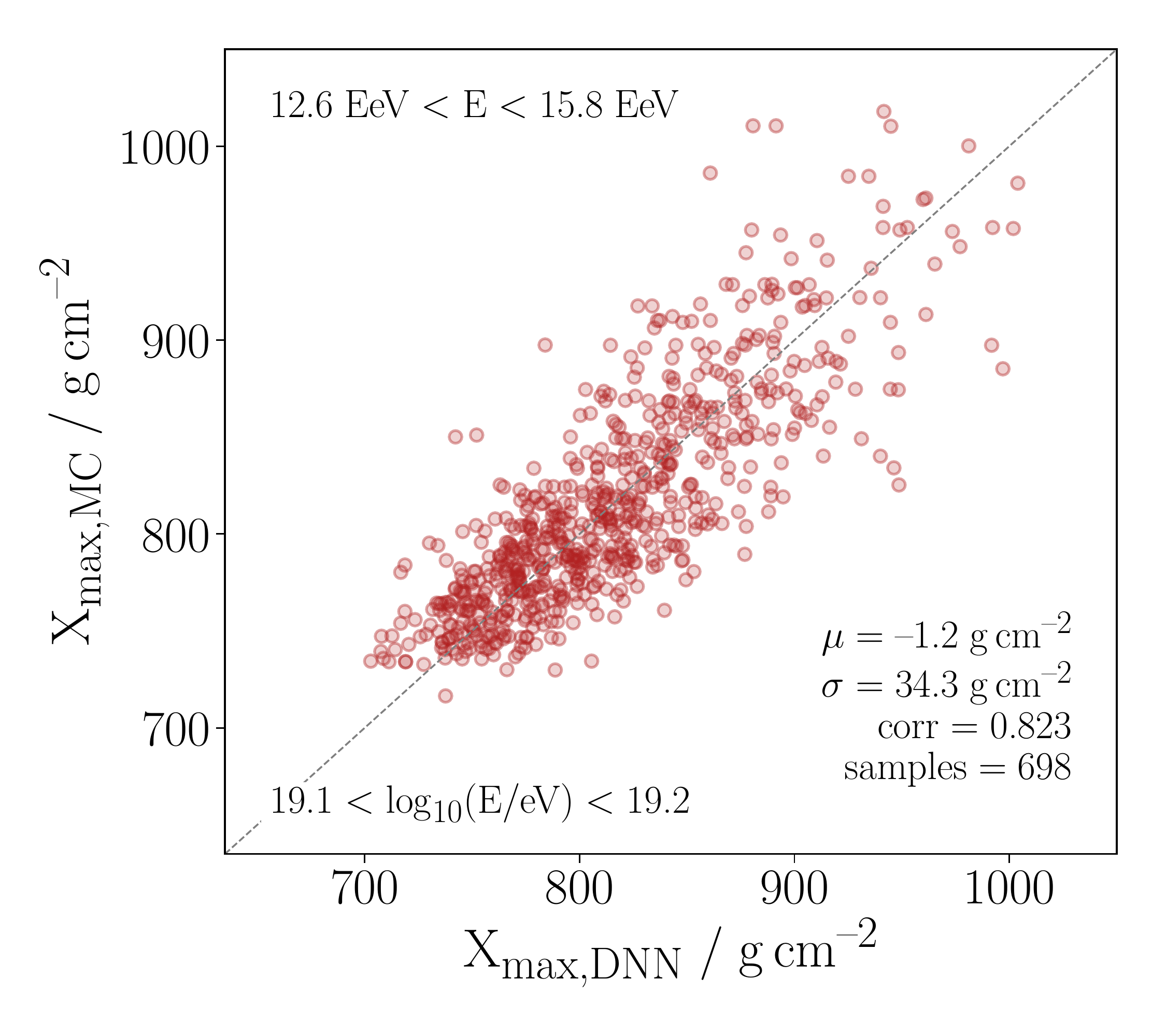}
                \subcaption{Proton}
                \label{fig:proton_corr}
            \end{centering}
        \end{subfigure}    
        \begin{subfigure}[b]{0.495\textwidth}
            \begin{centering}
                \includegraphics[width=0.99\textwidth, trim={0 0.45cm 0 0}, clip]{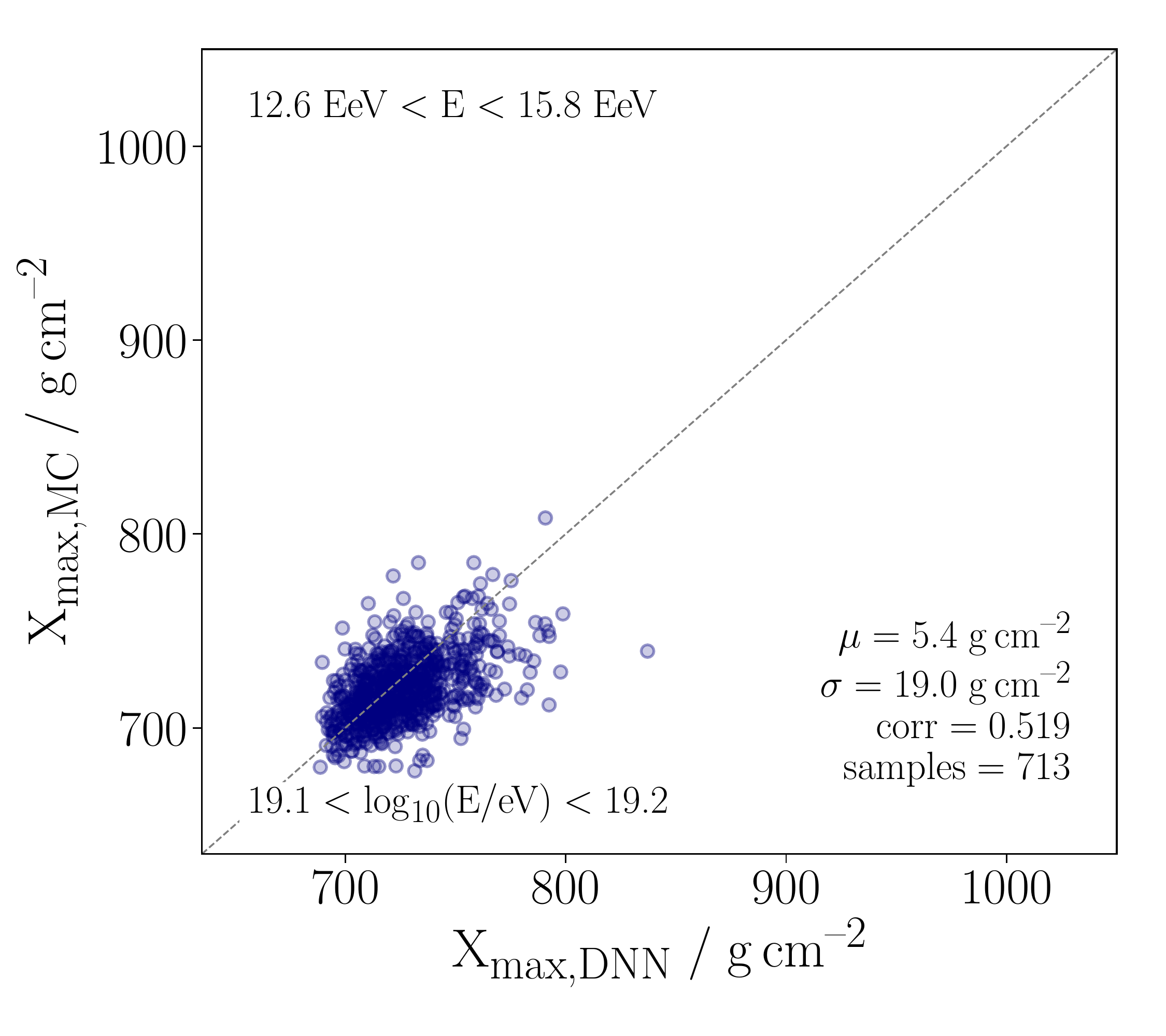}
                \subcaption{Iron}
                \label{fig:iron_corr}
            \end{centering}
        \end{subfigure}
    \caption{Correlation between \xmax reconstructed using the deep neural network and the true values simulated using EPOS-LHC in the energy range $19.1 < \log_{10}(E/\mathrm{eV}) < 19.2$ for proton (a) and iron (b) showers.}
    \label{fig:epos_corr}
    \end{centering}
\end{figure*}

\section{Performance on simulations}

In this section we study the performance of the deep neural network on the simulated data. Although none of the hadronic interaction models included in the simulations describe the measurements correctly, especially with respect to the muon deficit compared to data~\cite{muon_deficit}, we can at least investigate differences arising in the \xmax reconstruction when using different hadronic interaction models that predict different muon numbers. As latest results~\cite{icrc19_masscomp} indicate that EPOS-LHC is able to describe measurements of \xmax better than \qgsII, in this work the network was trained using the EPOS-LHC interaction model exclusively.

In the following we first investigate the results of the deep neural network when reconstructing \xmax of events also simulated with the EPOS-LHC model. Next, to study the influence when modifying the hadronic interaction model, we evaluate the neural network on other hadronic interaction models than those used for the training. We use simulations based on \qgsII and \sibyll as crosscheck in the following. We further tested the inverse setting, not presented in this work, observing that our findings are equivalent when flipping the sign of the biases.
Subsequently, we discuss the combined energy and zenith dependency of the \xmax reconstruction of the network, indicating the phase space region that can be used for a high quality reconstruction.
Finally, we investigate the reconstructed \xmax distributions.

\begin{figure*}[tb!]
    \begin{centering}
        \begin{subfigure}[b]{0.495\textwidth}
            \begin{centering}
                \includegraphics[width=0.99\textwidth, trim={0 0.45cm 0 0}, clip]{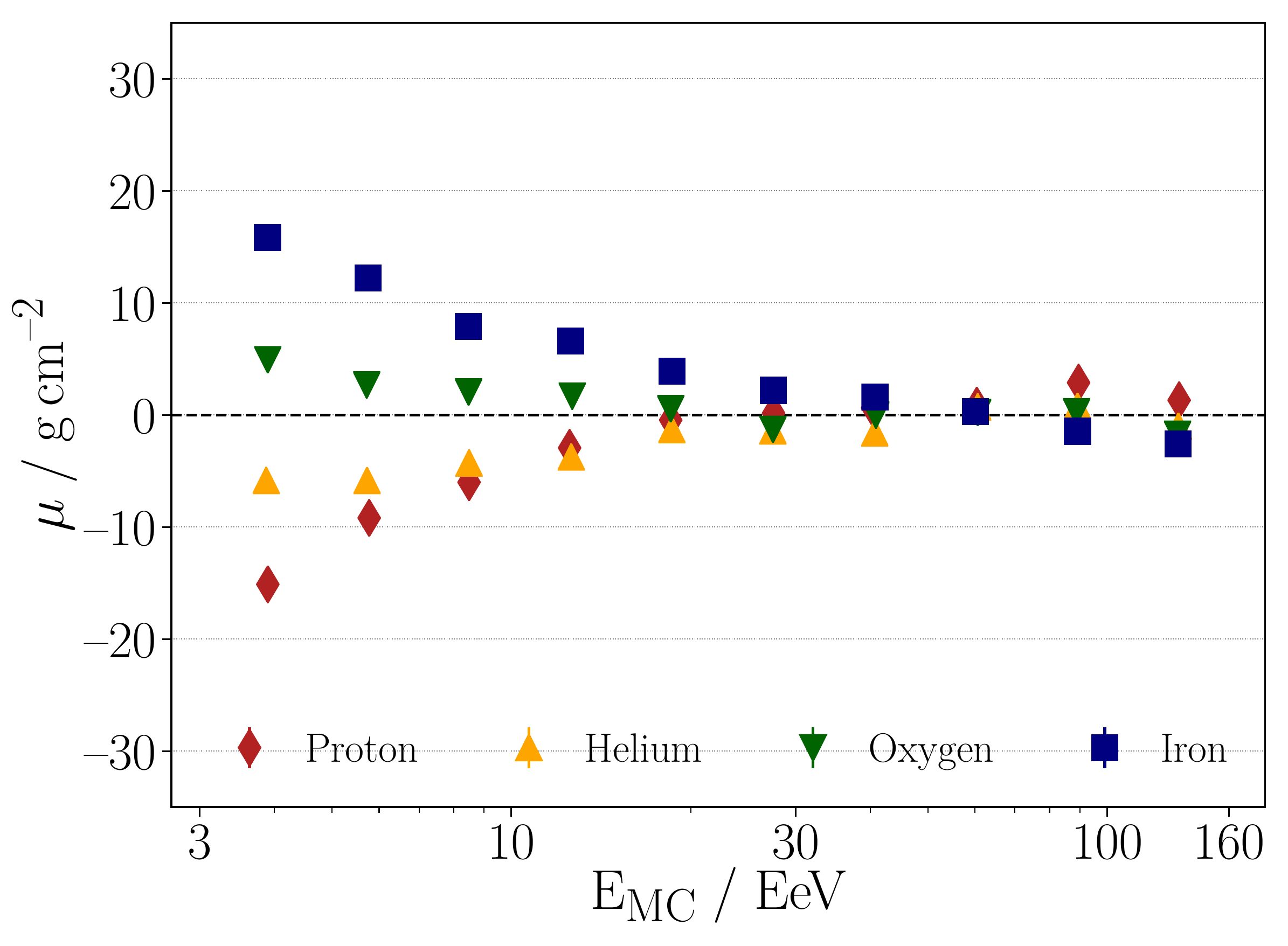}
                \subcaption{\xmax reconstruction bias}
                \label{fig:EPOS_energy_mu}
            \end{centering}
        \end{subfigure}
        \begin{subfigure}[b]{0.495\textwidth}
            \begin{centering}
                \includegraphics[width=0.99\textwidth, trim={0 0.45cm 0 0}, clip]{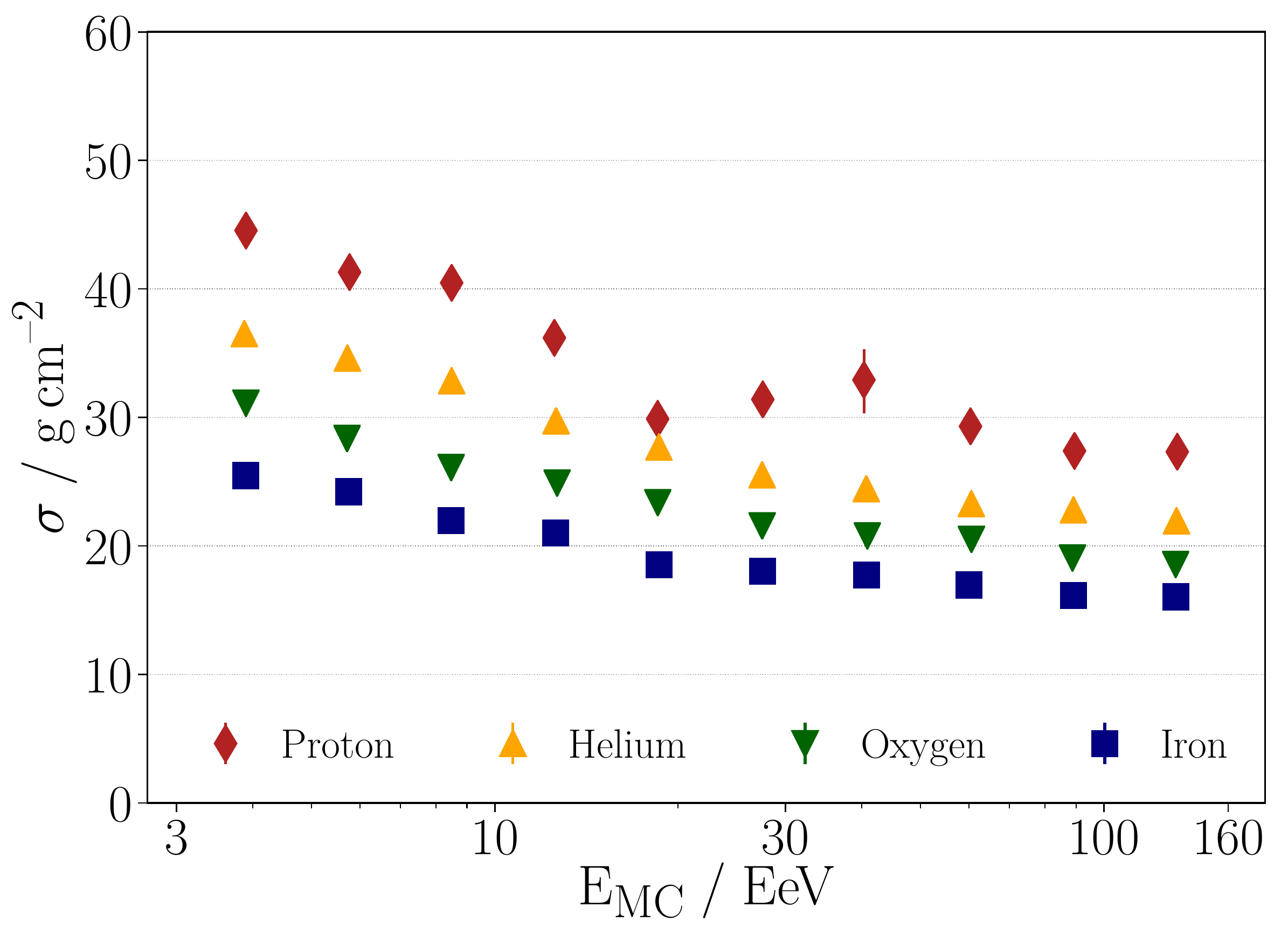}
                \subcaption{\xmax resolution}
                \label{fig:EPOS_energy_sigma}
            \end{centering}
        \end{subfigure}

    \caption{EPOS-LHC simulation study: Energy-dependent (a) bias and (b) resolution of the \xmax reconstruction \textit{evaluated} on EPOS-LHC showers with the deep neural network also \textit{trained} using EPOS-LHC showers. Different colors indicate different primaries.
}
    \label{fig:EPOS}
    \end{centering}
\end{figure*}

\subsection{Training and evaluation of the network using EPOS-LHC simulated events}
In the following we show the results of the \xmax reconstruction with the deep neural network as trained on simulations based on the EPOS-LHC interaction model. The events for evaluating the network originate from the  separated test set and were also simulated using the EPOS-LHC model.
In Fig.~\ref{fig:epos_corr} the event-by-event correlation of the reconstructed and true \xmax is shown for a benchmark bin. The Pearson correlation coefficient is above $0.5$ for all elements.
Comparing Fig.~\ref{fig:iron_corr} and Fig.~\ref{fig:proton_corr} implies that element-specific shower-to-shower fluctuations are taken into account in the deep-learning based reconstruction. This results in very different \xmax distributions for proton and iron showers reconstructed by the DNN. 

In Fig.~\ref{fig:EPOS_energy_mu} we show the \xmax reconstruction bias $\mu=\langle X_{\mathrm{max, DNN}} - X_{\mathrm{max, MC}}\rangle$ as a function of the cosmic ray energy. Above $10$~EeV the bias is below $\pm 9$~\gcm independent of the composition of the simulated event set used for evaluation. The statistical errors obtained via bootstrapping are mostly hidden by the markers due to the large statistics of the test set ($50{,}000$ events). Thus, the expected precision in the determination of the first moment of \xmax at the highest energies is excellent if the data were to look like the EPOS-LHC simulated data.

\begin{figure*}[t!]
    \begin{centering}
        \begin{subfigure}[b]{0.495\textwidth}
            \begin{centering}
                \includegraphics[width=0.99\textwidth, trim={0 0.45cm 0 0}, clip]{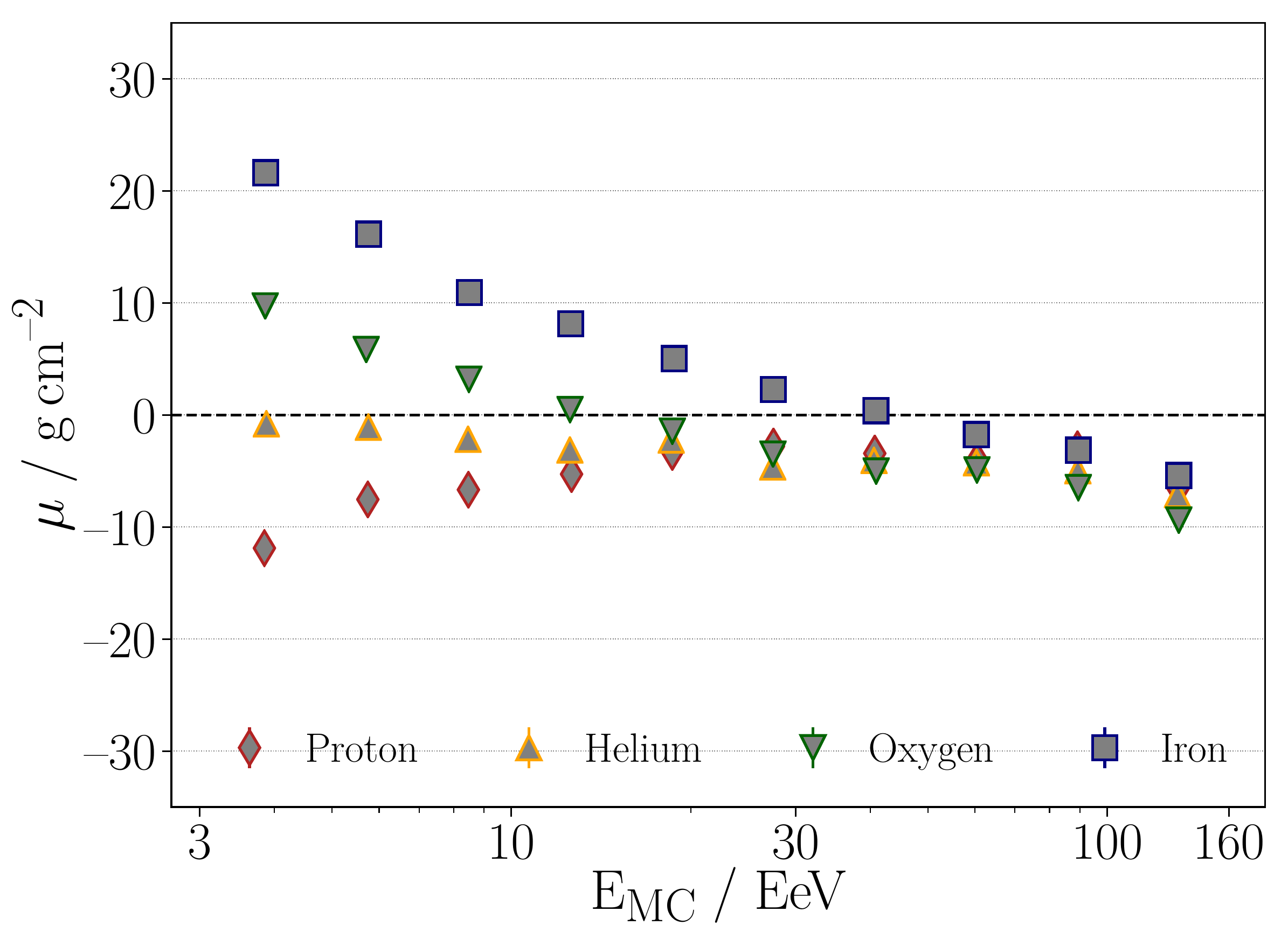}
                \subcaption{\qgsII: \xmax bias}
                \label{fig:QGS_energy_mu}
            \end{centering}
        \end{subfigure}
        \begin{subfigure}[b]{0.495\textwidth}
            \begin{centering}
                \includegraphics[width=0.99\textwidth, trim={0 0.45cm 0 0}, clip]{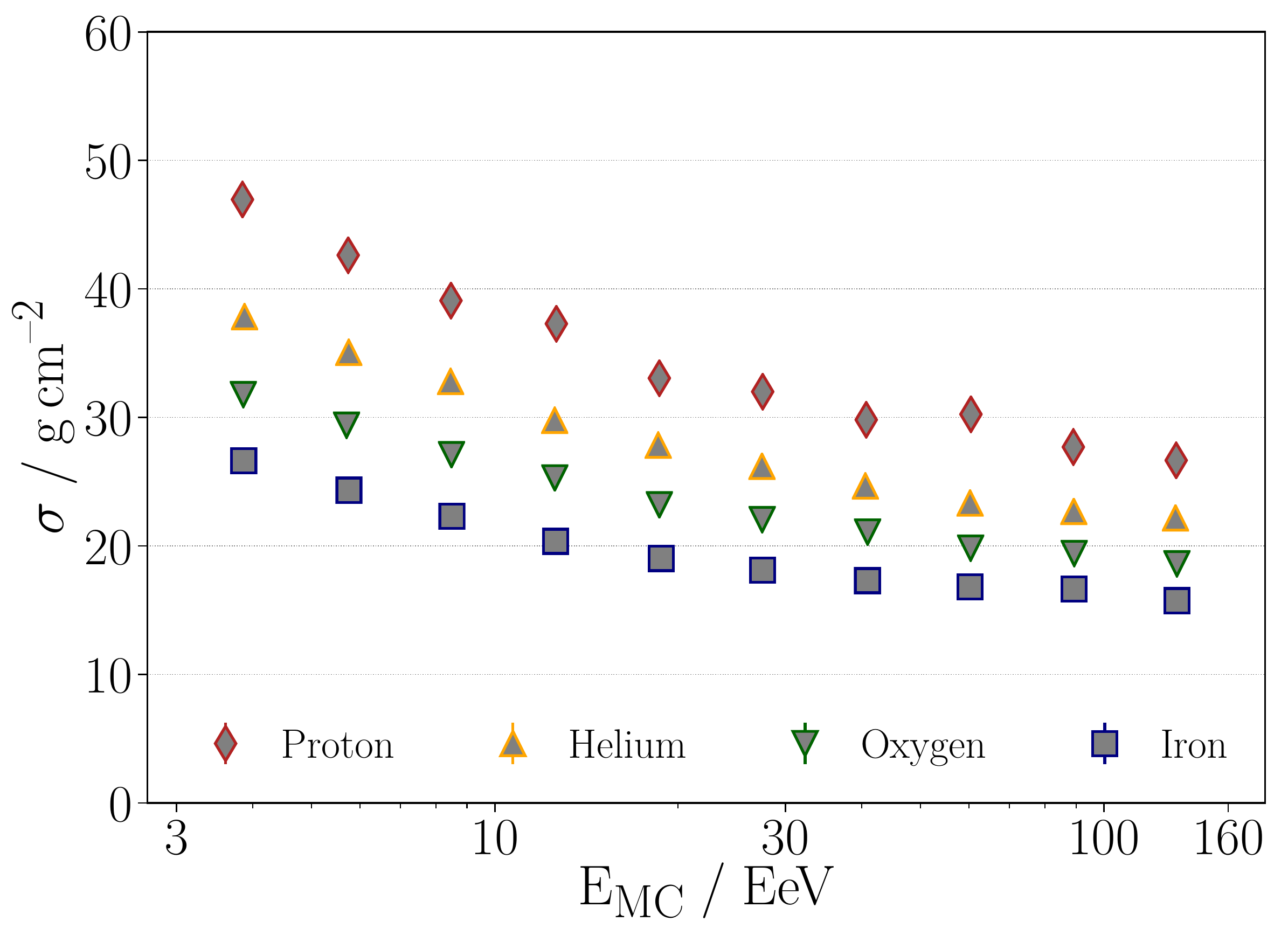}
                \subcaption{\qgsII: \xmax resolution}
                \label{fig:QGS_energy_sigma}
            \end{centering}
        \end{subfigure}
        \begin{subfigure}[b]{0.495\textwidth}
            \begin{centering}
                \vspace{0.4cm}
                \includegraphics[width=0.99\textwidth, trim={0 0.45cm 0 0}, clip]{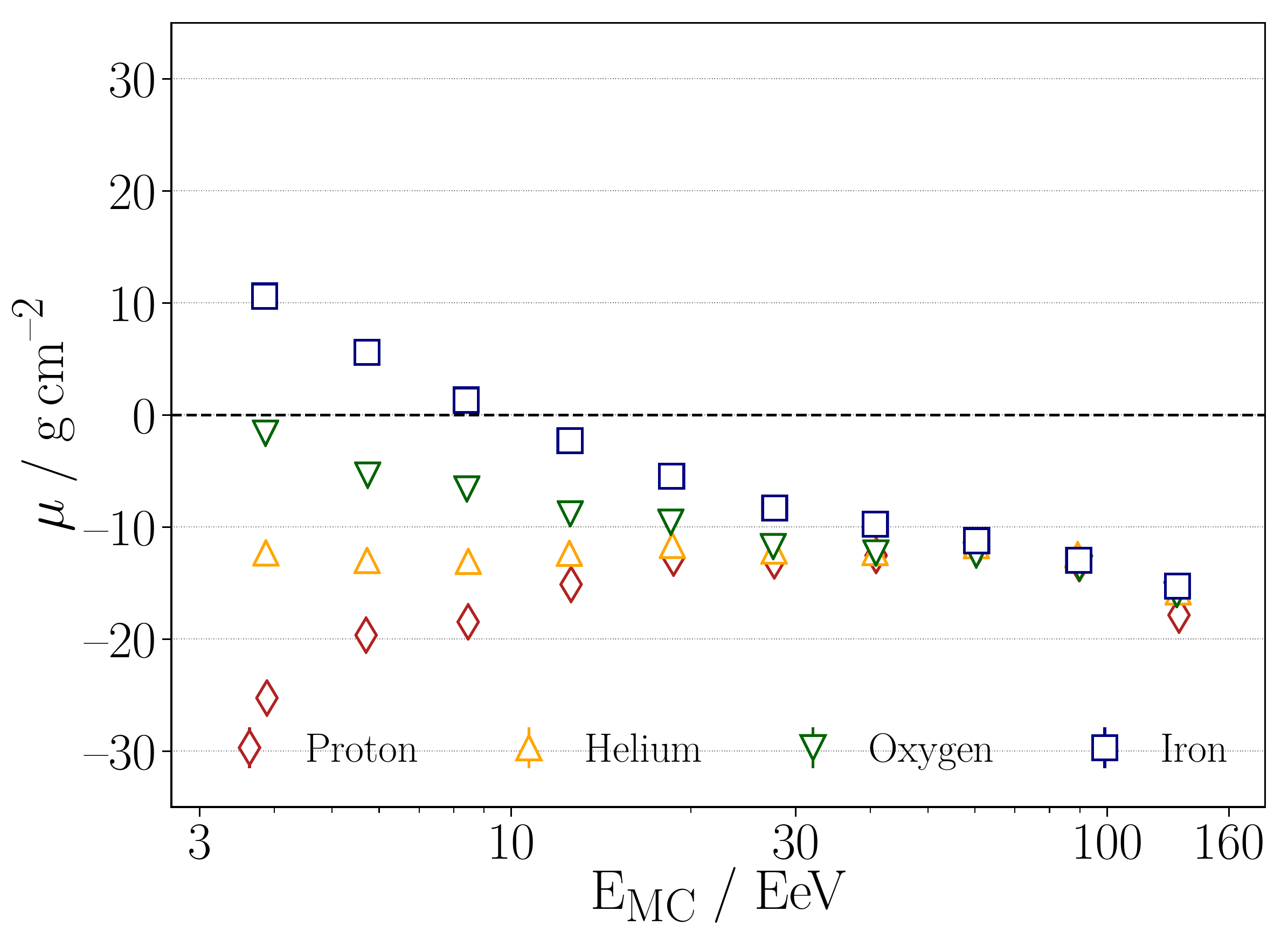}
                \subcaption{\sibyll: \xmax bias}
                \label{fig:SIB_energy_mu}
            \end{centering}
        \end{subfigure}
        \begin{subfigure}[b]{0.495\textwidth}
            \begin{centering}
                \includegraphics[width=0.99\textwidth, trim={0 0.45cm 0 0}, clip]{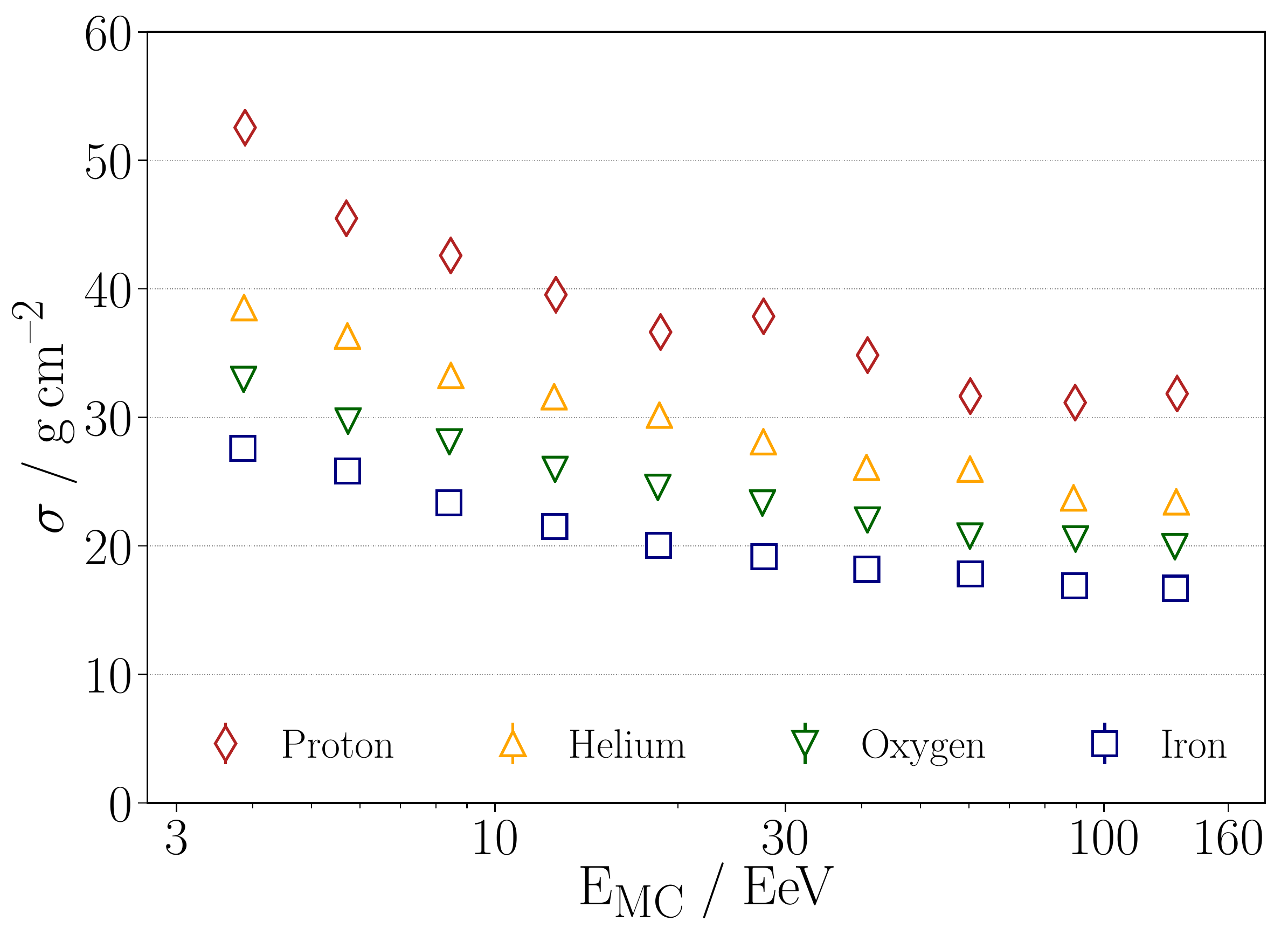}
                \subcaption{\sibyll: \xmax resolution}
                \label{fig:SIB_energy_sigma}
            \end{centering}
        \end{subfigure}
    \caption{Energy-dependent bias (left) and resolution (right) of the \xmax reconstruction \textit{evaluated} on simulated showers using the hadronic interaction models \qgsII (top) and \sibyll (bottom). The network was \textit{trained} using EPOS-LHC showers. Different colors indicate different primaries.}
    \label{fig:QGS_SIB}
    \end{centering}
\end{figure*}

In Fig.~\ref{fig:EPOS_energy_sigma} we show the event-by-event resolution $\sigma$ of the \xmax reconstruction as a function of the cosmic ray energy. The resolution improves with increasing energy, but exhibits a clear composition dependency.
This dependency is expected as showers initiated by lighter nuclei exhibit larger shower-to-shower fluctuations.
At $10$~EeV the \xmax resolution for proton-induced showers reaches $38$~\gcm which is below the \xmax fluctuations of $40$~\gcm as measured by the Pierre Auger Observatory~\cite{icrc19_masscomp}. For iron induced showers the \xmax resolution at $10$~EeV is at the level of $20$~\gcm only.

\subsection{Evaluation of the network on \qgsII and \sibyll simulated events}

In order to investigate the dependency of the deep neural network on the hadronic interaction model used for simulations, we evaluate our network trained on showers simulated using the EPOS-LHC model on simulations using \qgsII and \sibyll.
In Fig.~\ref{fig:QGS_SIB} we show the event-by-event resolution $\sigma$ and bias $\mu$ of the \xmax reconstruction as a function of the cosmic ray energy.
In Fig.~\ref{fig:QGS_energy_mu} we show the reconstruction bias $\mu$ as a function of the cosmic ray energy for \qgsII. Above $10$~EeV the bias is below $\pm 10$~\gcm, however with a shift of approximately $-5$~\gcm at larger energies. The shift observed for simulations using \sibyll (see Fig.~\ref{fig:SIB_energy_mu}) is slightly larger and amounts to roughly $-15$~\gcm. We assign this effect to the hadronic interaction model as it was not visible in Fig.~\ref{fig:EPOS_energy_mu}.
Besides the total shift, a slight energy-dependency can be observed when averaging among the compositions of \sibyll and \qgsII.

Tracing back the differences to individual characteristics of the hadronic interaction models is highly complex since the method is based on time-dependent signals. Therefore, not only the abundance of the individual shower components but also their respective time-dependent shower development needs to be considered.

The resolution of the network consistently improves with increasing energy, but exhibits a clear composition dependency which is the same for both interaction models.
At $10$~EeV the \xmax resolution for proton-induced showers reaches $38$~\gcm and for iron showers $20$~\gcm as visible in Fig.~\ref{fig:QGS_energy_sigma} and Fig.~\ref{fig:SIB_energy_sigma}. These values are close to the result shown in Fig.~\ref{fig:EPOS_energy_sigma}, so we conclude that the resolution of \xmax is almost independent of the hadronic interaction model.

The element-wise event-by-event correlations are very similar for all elements and energies.
This is not surprising since the correlation is connected to the resolution which was found to be independent of the hadronic interaction model.

\subsection{Zenith dependency of the reconstruction}
\begin{figure*}[t!]
    \begin{centering}
        \begin{subfigure}[b]{0.495\textwidth}
            \begin{centering}
                \includegraphics[width=0.99\textwidth, trim={0 0.55cm 0 0}, clip]{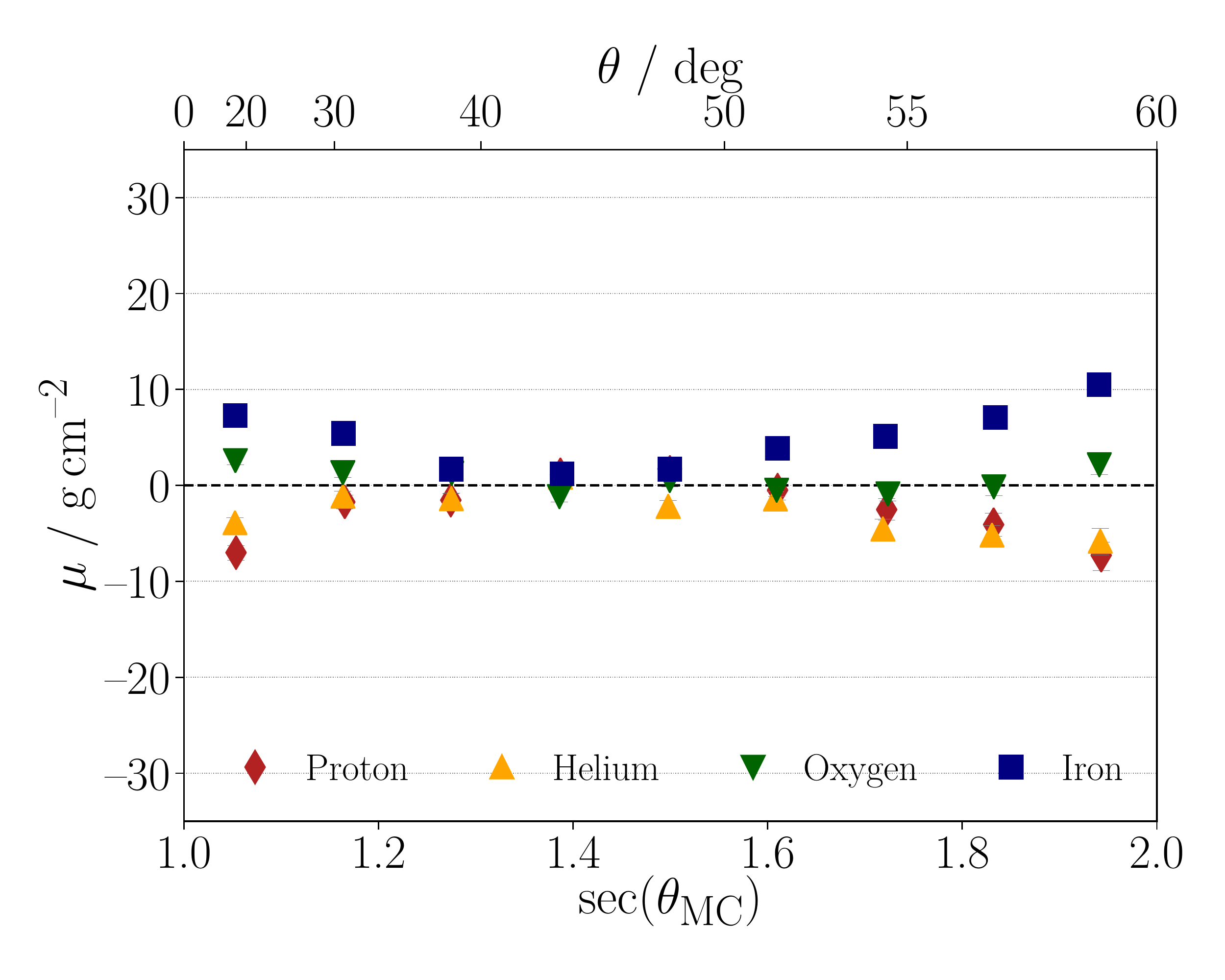}
                \subcaption{}
                \label{fig:EPOS_zenith_mu}
            \end{centering}
        \end{subfigure}
        \begin{subfigure}[b]{0.495\textwidth} 
            \begin{centering}
                \includegraphics[width=0.99\textwidth, trim={0 0.55cm 0 0}, clip]{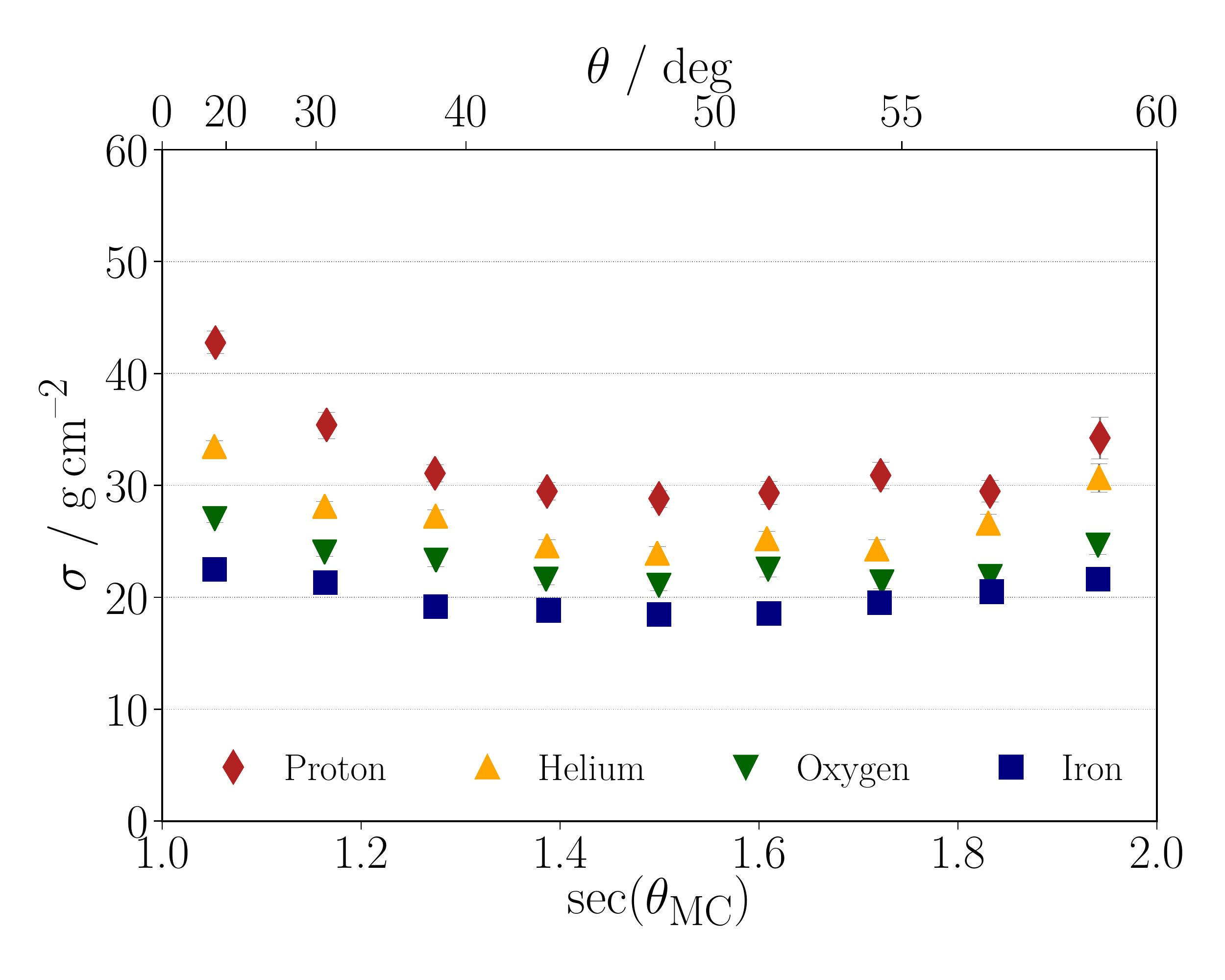}
                \subcaption{}
                \label{fig:EPOS_zenith_sigma}
            \end{centering}
        \end{subfigure}
    \caption{EPOS-LHC simulation study: Zenith-dependent (a) bias and (b) resolution of the \xmax reconstruction \textit{evaluated} on EPOS-LHC showers with the deep neural network also \textit{trained} using EPOS-LHC showers. Different colors indicate different primaries.}
    \label{fig:EPOS_zenith}
    \end{centering}
\end{figure*}

To investigate the zenith dependency of the reconstruction, the evaluation of the neural network on air showers simulated with EPOS-LHC is presented.
For the zenith angle in Fig.~\ref{fig:EPOS_zenith_mu}, we observe a moderately varying, composition-dependent reconstruction bias of up to $\pm 10$~\gcm at $\theta=60^{\circ}$. Also the \xmax resolution in Fig.~\ref{fig:EPOS_zenith_sigma} exhibits moderate variations with the zenith angle. For proton-induced showers they are around $30$~\gcm, and for iron around $20$~\gcm. For all elements a particular good reconstruction is visible at $\theta \approx 45^{\circ}$ and a worsening for angles $\theta > 55^{\circ}$ and  $\theta < 20^{\circ}$. The reason for this behavior is the concurrence of larger footprints and increasing absorption effects of the atmosphere for larger zenith angles, leading to more triggered WCDs but reducing the number of particles measured by an individual detector station. Further, for very vertical showers the shower maximum can be close to the ground or even underground which makes the reconstruction more difficult.

\begin{figure*}[t!]
    \begin{center}
        \begin{subfigure}[b]{0.495\textwidth} 
            \begin{centering}
                \includegraphics[width=0.99\textwidth, trim={0.6cm 0 0.5cm 0.50cm}, clip]{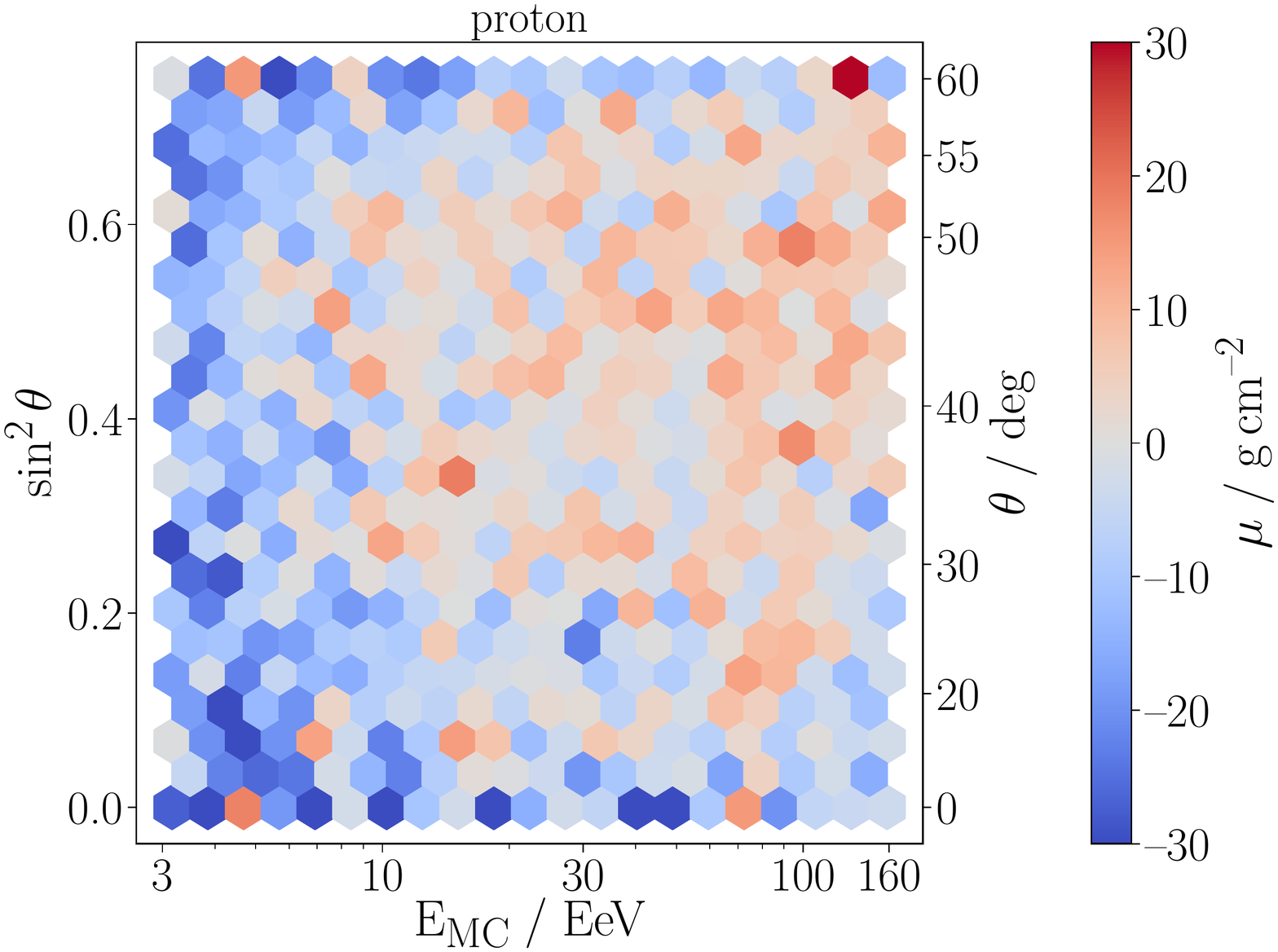}
                \subcaption{}
                \label{fig:zenith_energy_bias_proton}
            \end{centering}
        \end{subfigure}
        \begin{subfigure}[b]{0.495\textwidth} 
            \begin{centering}
                \includegraphics[width=0.99\textwidth, trim={0.6cm 0 0.5cm 0.50cm}, clip]{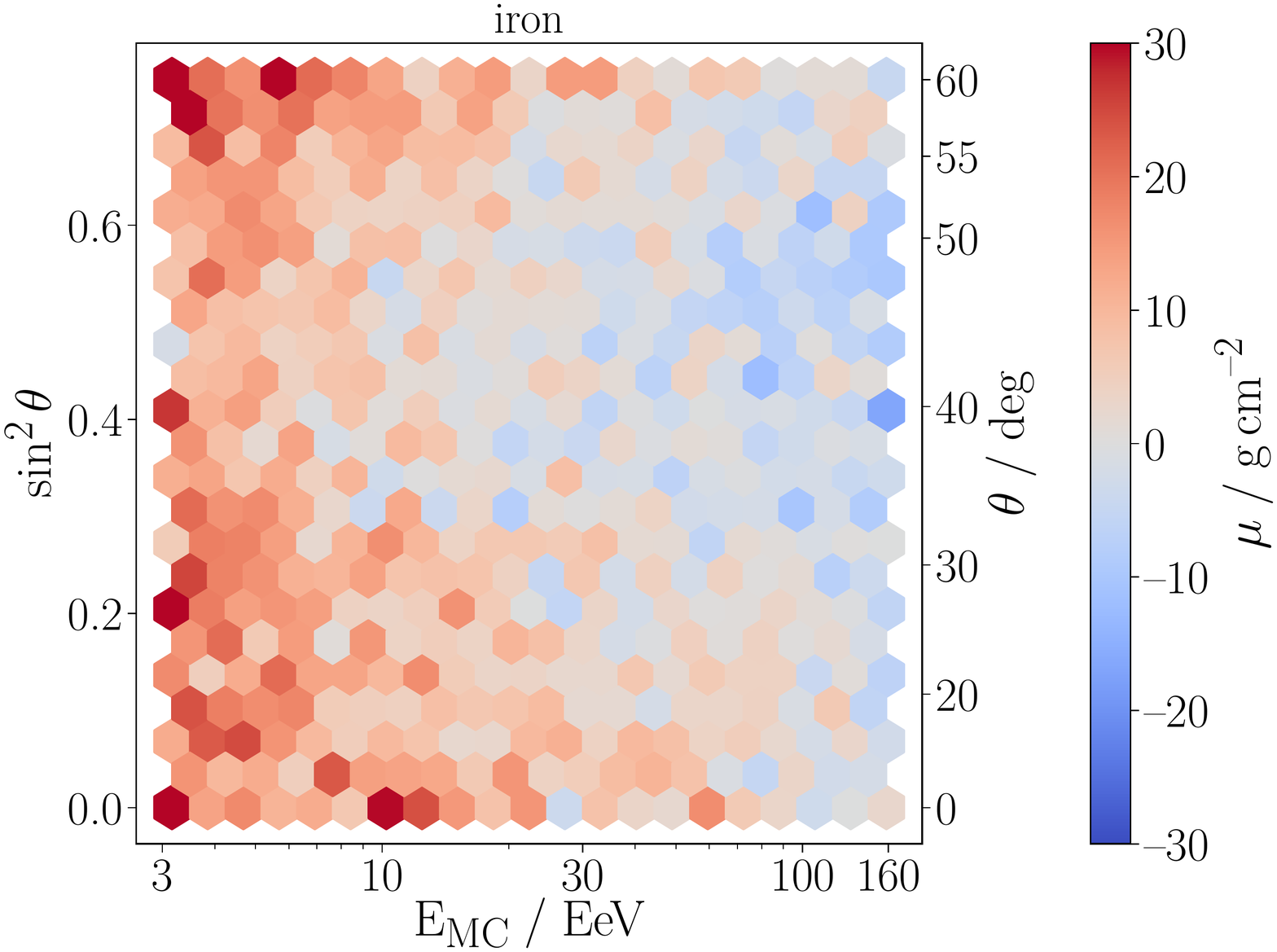}
                \subcaption{}
                \label{fig:zenith_energy_bias_iron}
            \end{centering}
        \end{subfigure}
        \caption{Energy- and zenith-dependent bias of the deep neural network trained and evaluated using EPOS-LHC for proton (a) and iron (b) showers.}
        \label{fig:zenith_energy_bias}
    \end{center}
\end{figure*}

Note that the results shown in Fig.~\ref{fig:EPOS_zenith} are dominated by events with low energies. To verify a phase-space region in energy and zenith angle for high-quality reconstruction of \xmax we present in Fig.~\ref{fig:zenith_energy_bias} the zenith-energy dependency of the \xmax bias separately for protons and iron nuclei which show the largest bias. To obtain the same number of events in each bin, we plot $\sin^2\theta$ and use logarithmic bins in energy.
It is evident that a high-quality phase space at moderate zenith angles exists already at smaller energies, its size increases with energy and allows for reconstructions with a small bias. Outside this region for very vertical and horizontal showers, the zenith bias is particularly apparent for proton primaries, but remains mostly in the order of $15$~\gcm per energy bin over the complete energy range. For iron nuclei the bias is even smaller. In addition, vertical proton showers with high energies show an increased bias as for such shower geometries shower maxima below ground are more likely.
These found dependencies can also be observed with showers simulated using \qgsII and \sibyll.

Overall the reconstruction of \xmax with respect to its bias and resolution performs well on simulations. Above cosmic ray energies of $10$~EeV, the expected \xmax reconstruction bias is below $\pm 10$~\gcm and the \xmax resolution becomes better than $35$~\gcm even for the lightest particles. When investigating events with saturated stations, we do not find noteworthy differences in the network performance.

Although we observe slight differences caused by using different hadronic interaction models, the network reconstruction exhibits only minor dependencies on the interaction models.

\subsection{Distribution of the reconstructed shower maxima}

For estimating the cosmic-ray composition the correct shape of the \xmax distribution is essential.
Due to the fact that the DNN is trained using the mean squared error, the predicted distributions are not broadened by the resolution of the method, but tend to be truncated. This bias towards the mean of the true \xmax distribution depends on the performance and creates a dependency on \xmax. Consequently, the quality of the estimator cannot be determined by the resolution only and is therefore examined in detail.
As a result, even if the resolution for iron showers is in the range of the shower-to-shower fluctuations, a distribution with correct shape can be predicted if the estimator is sufficiently precise. Precise means in this case that the DNN assigns proton-like values of \xmax to proton-like showers and iron-like values of \xmax to iron-like showers etc.

\begin{figure*}[t!]
    \begin{centering}
        \begin{subfigure}[b]{0.495\textwidth}
            \begin{centering}
                \includegraphics[width=0.99\textwidth]{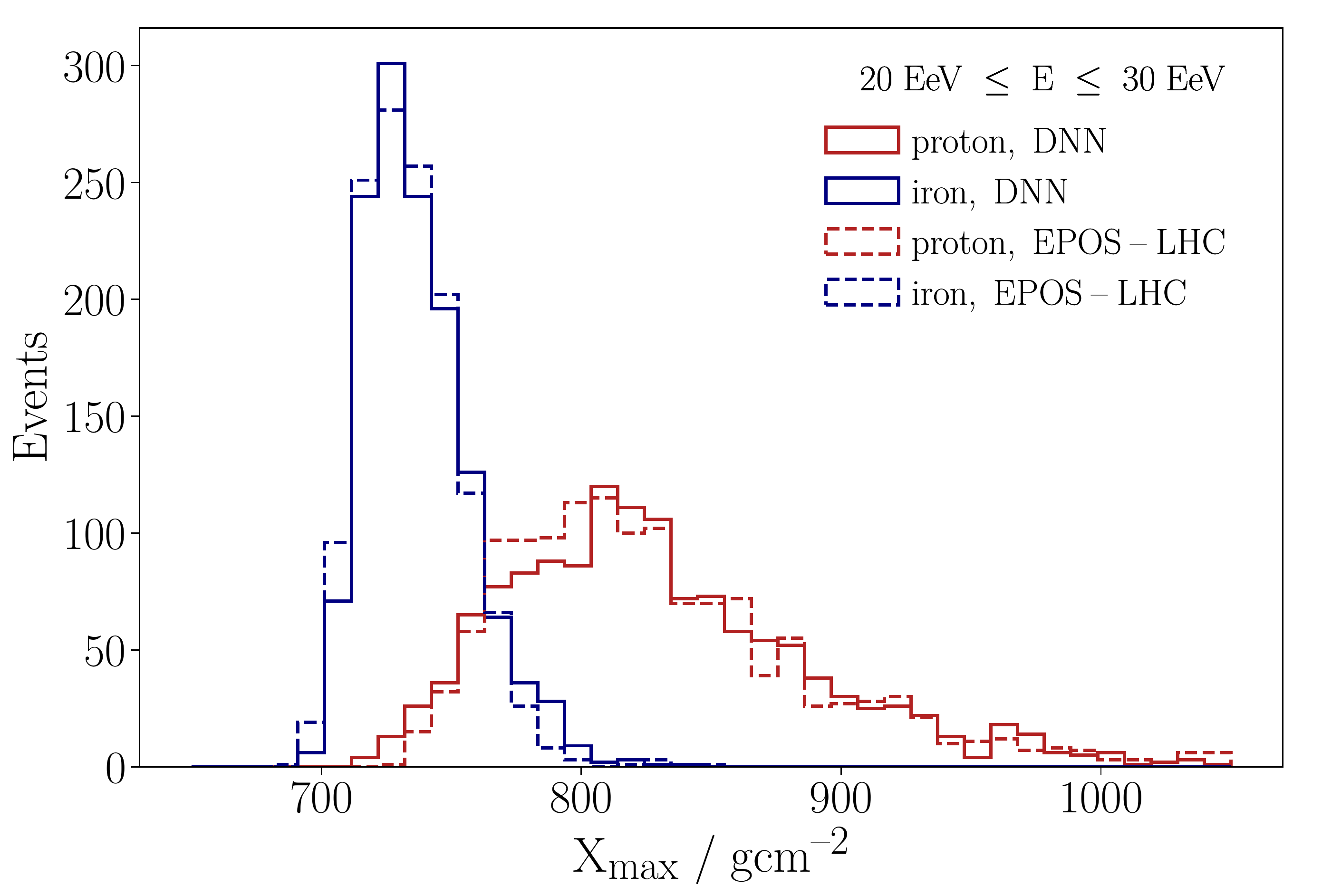}
                \subcaption{}
                \label{fig:epos_moment}
            \end{centering}
        \end{subfigure}
        \begin{subfigure}[b]{0.495\textwidth}
            \begin{centering}
                \includegraphics[width=0.99\textwidth]{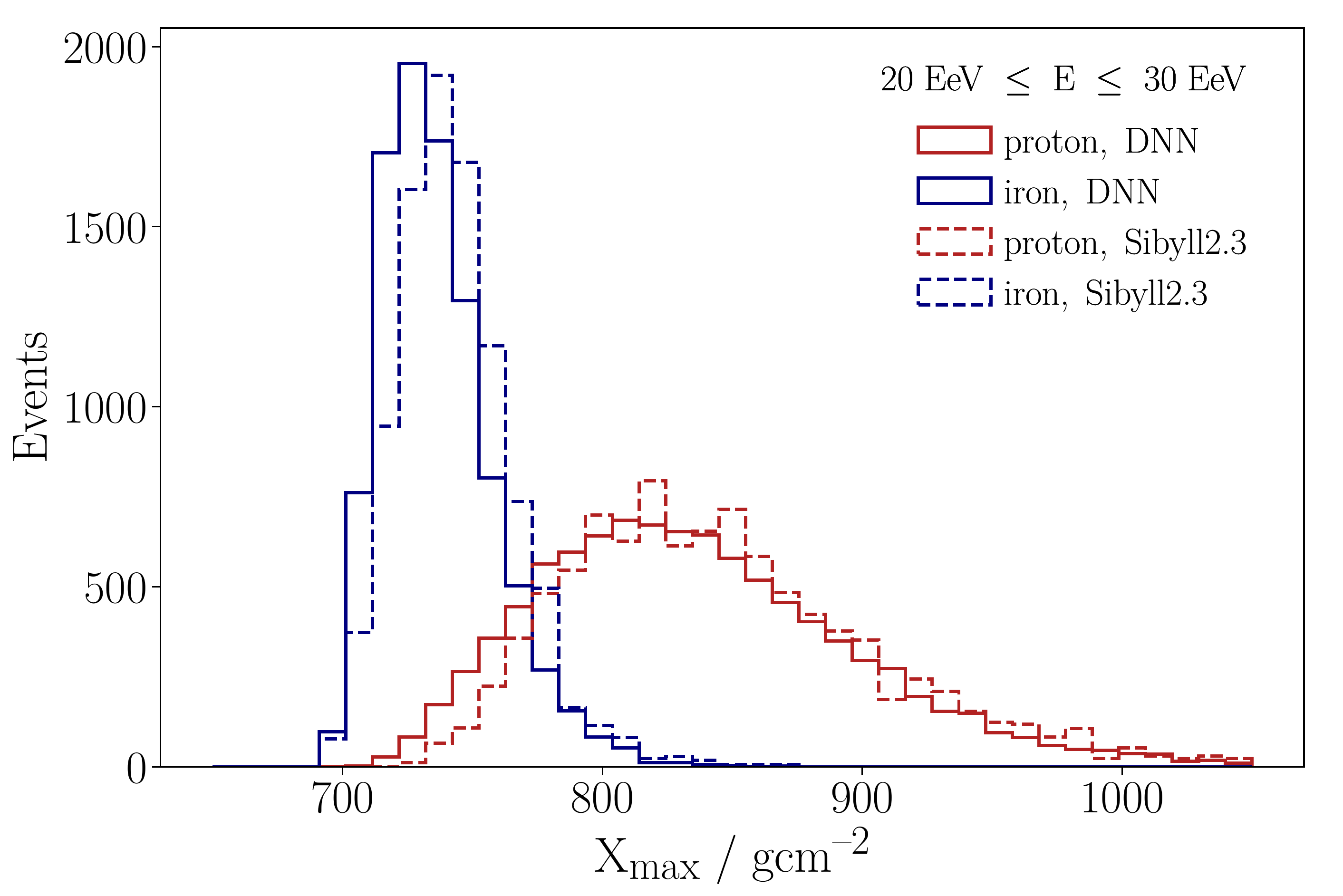}
                \subcaption{}
                \label{fig:sib_moment}
            \end{centering}
        \end{subfigure}
    \caption{Comparison of the \xmax distributions as reconstructed with the network (solid histograms) for proton (red) and iron (blue) showers and the simulated distribution (dashed histograms) for (a) EPOS-LHC and (b) \sibyll for energies between $20$ and $30$~EeV.}
    \label{fig:gumbel}
    \end{centering}
\end{figure*}

To investigate the reconstructed \xmax distribution of the network for different energies we show in Fig.~\ref{fig:gumbel} the results of the \xmax reconstruction with the deep neural network for two example bins.
In Fig.~\ref{fig:epos_moment} the reconstructed distribution at $20-30$~EeV for showers simulated with EPOS-LHC is shown. 
The solid histograms represent the distribution of reconstructed \xmax for proton (red) and iron (blue) showers. The simulated \xmax values are denoted by dashed histograms. It is apparent that both, the proton distribution and the iron distribution, are in good agreement.
In Fig.~\ref{fig:sib_moment} the reconstructed \xmax distribution for showers simulated using \sibyll at the same energy is shown. Excepting a total bias of around $15$~\gcm which we expect from the observed hadronic-model bias (see Fig.~\ref{fig:SIB_energy_mu}) the simulated and reconstructed distributions match well.
The successful reconstruction of the \xmax distributions can be generalized for energies $\geq 10$~EeV. At lower energies the reconstruction biases at low and high zenith angles (compare Fig.~\ref{fig:zenith_energy_bias}) superimpose the physics fluctuations. A fiducial selection could potentially reduce this effect.

The observation that \xmax distributions similar to the true values are reproduced for all elements and hadronic interaction models (despite an absolute offset), implies that the neural network can be used to gain insights beyond the first moment of the distribution. Beyond the measurement of $\sigma(\xmax)$, this opens up possibilities for in-depth analyses of the mass-fractions of UHECRs~\cite{fd_xmaxI}.

\section{Application to hybrid data}

In this section we evaluate the performance of the deep neural network on events which include measurements of the surface detector and the fluorescence detector. First, we correct for long-term aging effects of the WCDs, which are moderate over the many years of operation, but are still noticeable in the network prediction. Then we calibrate the absolute value of the reconstructed \xmax using hybrid measurements. Finally, we will determine the \xmax resolution to verify the potential for obtaining new information about the cosmic-ray composition at the highest energies from the first two moments of the \xmax distribution.

\subsection{Corrections for detector-aging effects}
\label{sec:aop}

\begin{figure*}[t!]
    \begin{center}
        \begin{subfigure}[b]{0.495\textwidth} 
            \begin{centering}
                \includegraphics[width=0.99\textwidth]{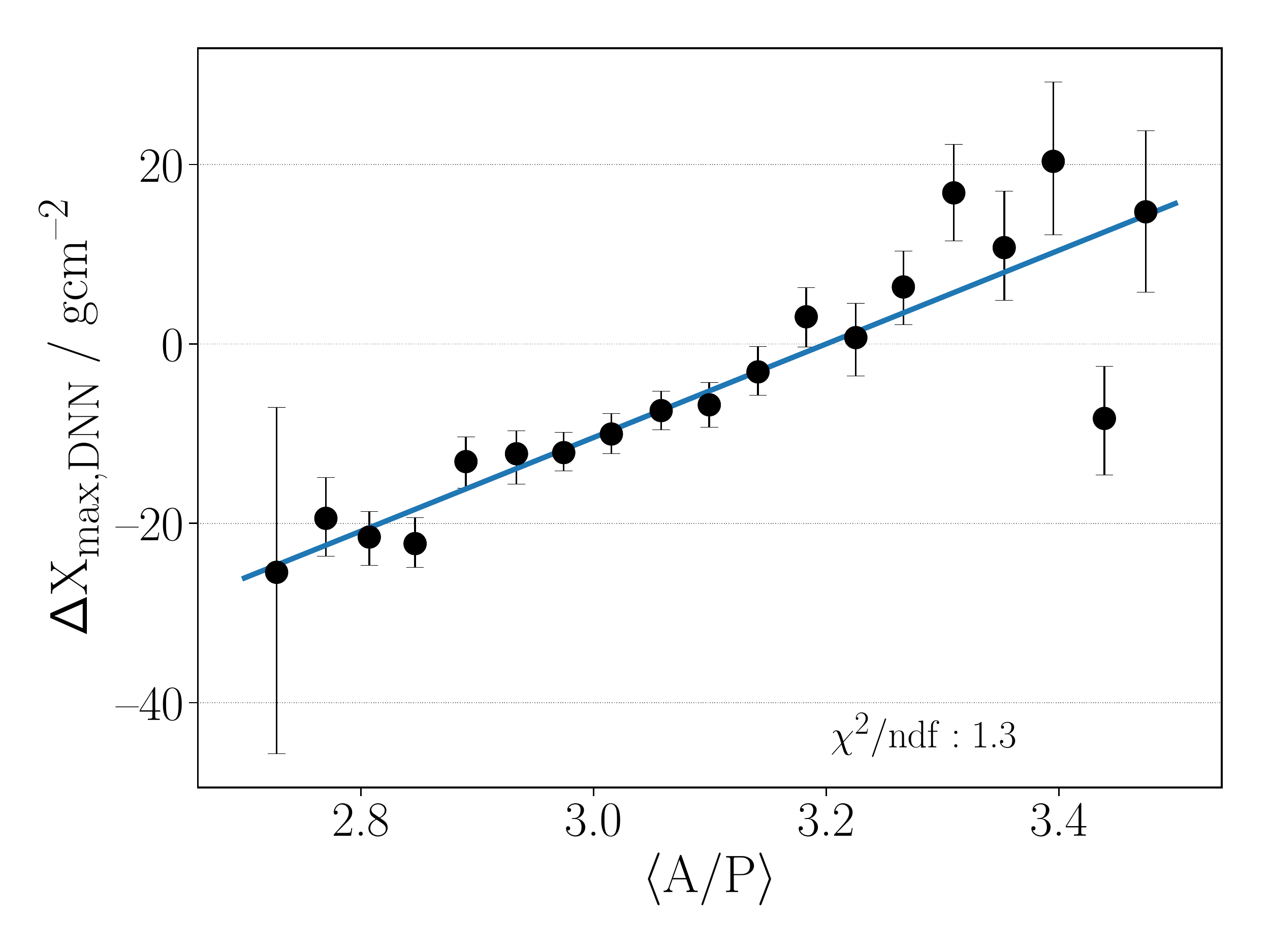}
                \subcaption{}
                \label{fig:aop_bias}
            \end{centering}
        \end{subfigure}
        \begin{subfigure}[b]{0.495\textwidth} 
            \begin{centering}
                \includegraphics[width=0.99\textwidth]{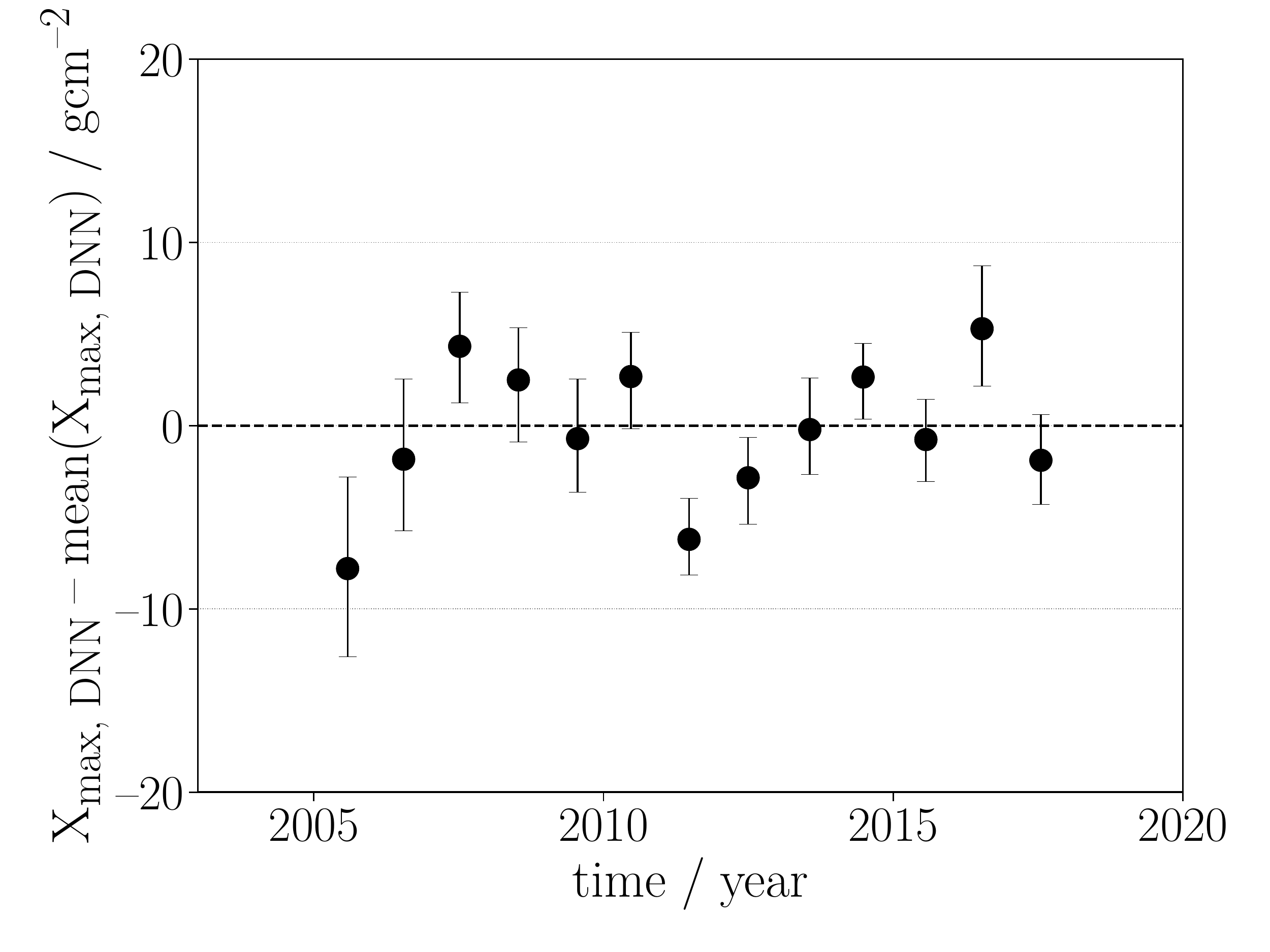}
                \subcaption{}
                \label{fig:after_calib}
            \end{centering}
        \end{subfigure}
        \caption{(a) Calibration of detector and aging effects. Describing the observed \xmax difference with respect to measured events with calibration parameters close to the simulation $(A/P)_\mathrm{sim}=3.2$, plotted versus the averaged area over peak ($\langle A/P \rangle $) variable. (b) Stable long-term dependency of the average \xmax after the $A/P$ calibration.}
        \label{fig:aop_}
    \end{center}
\end{figure*}

The exact shape of the signal traces has a decisive influence on the network predictions (see Fig.~\ref{fig:trace}). Besides the slightly varying response of each PMT (see section~\ref{augmentation}), over the long operating time of the observatory, aging effects in the signals of the WCDs occur as a combined effect of the water quality, the reflective Tyvek coating, the PMTs, and the electronics. The influence of these effects is continuously monitored by the local calibration using atmospheric muons that constantly pass through the detectors~\cite{trigger}. This allows to quantify measured signals in the unit of vertical equivalent muons (VEM). Every $60$~s the calibration parameters $Q_{\mathrm{VEM}}$ and $I_{\mathrm{VEM}}$ are determined using the signal pulses induced by atmospheric muons. Here, $I_{\mathrm{VEM}}$ corresponds to the pulse-height induced by a vertical traversing muon and $Q_{\mathrm{VEM}}$ to the integrated pulse (bins of $25$~ns).

To correct for slightly different pulse-shapes and effects of aging, the \emph{area over peak} (${A/P}$) variable has demonstrated its effectiveness~\cite{longterm}. It describes the ratio $\mathrm{A/P} = Q_{\mathrm{VEM}}/I_{\mathrm{VEM}}$, and hence is a measure of the pulse shape by quantifying the relation between the signal height and decay. Due to the FADC sampling rate of $40$~MHz, $A/P$ is given in units of $25$~ns. As first order approximation the area over peak is averaged for each event over all PMTs of triggered stations, which results to a characteristic and event-wise $\langle {A/P} \rangle$.

The gradual aging effects of the detectors are measurable in the detector monitoring and the predictions of the neural network. Since the beginning of the operation $\langle {A/P} \rangle$ decayed from roughly $3.2$ to $2.95$. During that time, the predicted $\langle X_{\mathrm{max}} \rangle$ of the DNN decayed by about $1~\gcm$ per year.
As in simulations all detectors show the same ${(A/P)}_\mathrm{sim} = 3.2$, we calibrate the predicted $X_{\mathrm{max}, A/P=3.2}$ of the network with respect to events with $\langle A/P \rangle$ close to the simulated values, where we expect a high-quality reconstruction.
The calibrated predictions give
\begin{equation}
X_{\mathrm{max}} = X_{\mathrm{max}, A/P=3.2} + b(\langle A/P \rangle)\; ,
\end{equation}
where $b$ is the observed difference. The difference shows a clear correlation with $\langle A/P \rangle$ (see Fig.~\ref{fig:aop_bias}) and is parameterized using:
$b(\langle A/P \rangle) = a \cdot (\langle A/P \rangle - 3.2)$,
with the fitted calibration constant $a=52.2 \pm 3.8~\gcm$.
This removes the influence of the aging effects of the water-Cherenkov detectors on the network prediction of \xmax as shown in Fig.~\ref{fig:after_calib}.
Due to the difference of the $A/P$-values in data and simulations, the average \xmax prediction is increased by roughly $9$~\gcm using the $A/P$ calibration.

\begin{figure}[t]
    \begin{center}
        \includegraphics[width=0.6\textwidth]{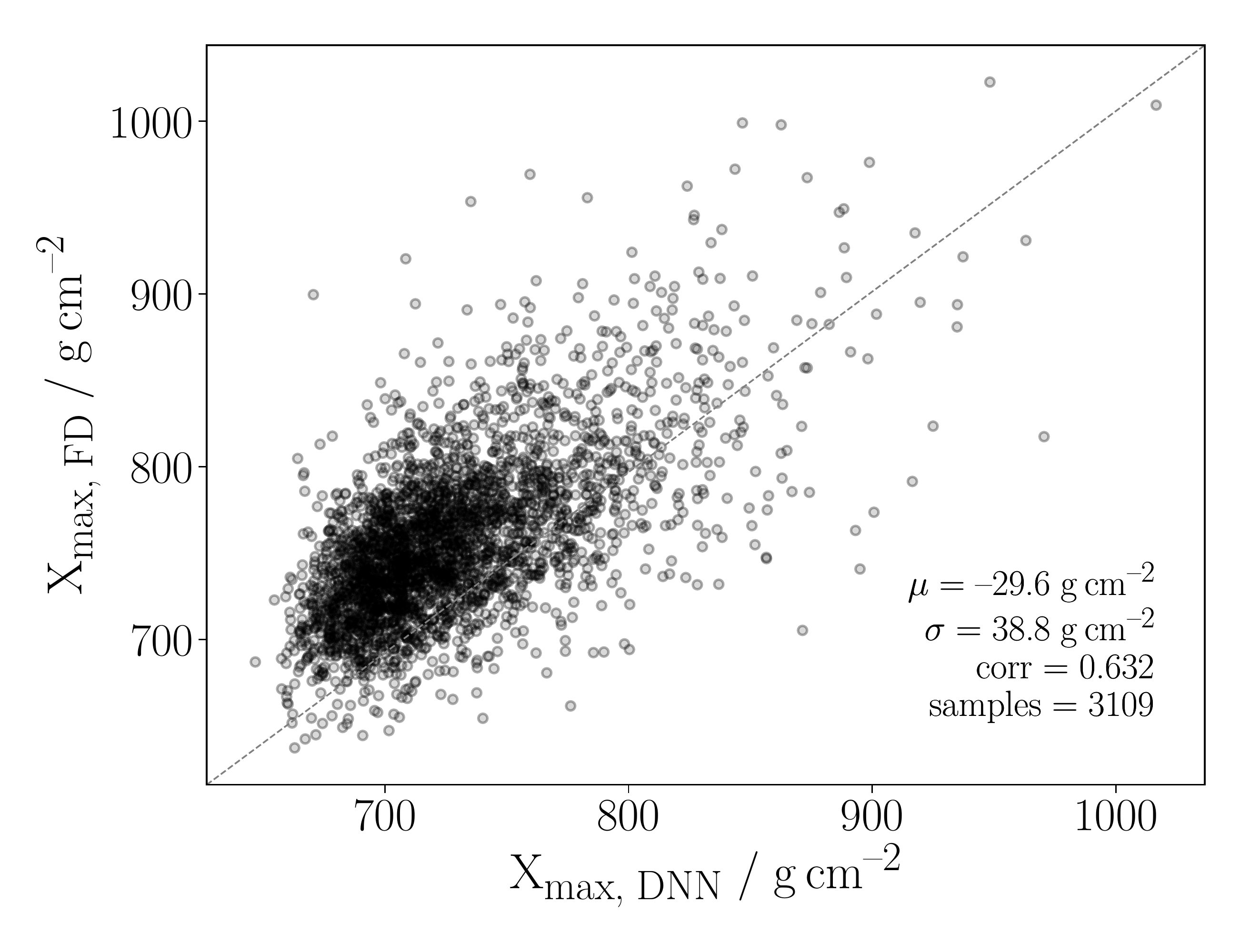}
        \caption{Event-by-event correlation of the \xmax reconstruction of the deep neural network and their counterparts reconstructed by the FD.}
        \label{fig:correlation}
    \end{center}
\end{figure}

\subsection{Evaluation and calibration of the first moment of the distribution of shower maxima}
To further quantify the reconstructions of the DNN and allow for an absolute calibration of \xmax we use hybrid events which offer high-quality fluorescence measurements. 
In Fig.~\ref{fig:correlation} the event-by-event correlation between the \xmax reconstruction of the DNN and the FD is shown.
The Pearson correlation coefficient amounts to $0.63$ and remains above $0.6$ even if \xmax is corrected for the elongation rate observed by the FD~\cite{fd_xmaxI}.

\begin{figure*}[t!]
    \begin{centering}
        \begin{subfigure}[b]{0.325\textwidth}
            \begin{centering}
                \includegraphics[width=0.98\textwidth, trim={0.95cm 0 0.5cm 0}, clip]{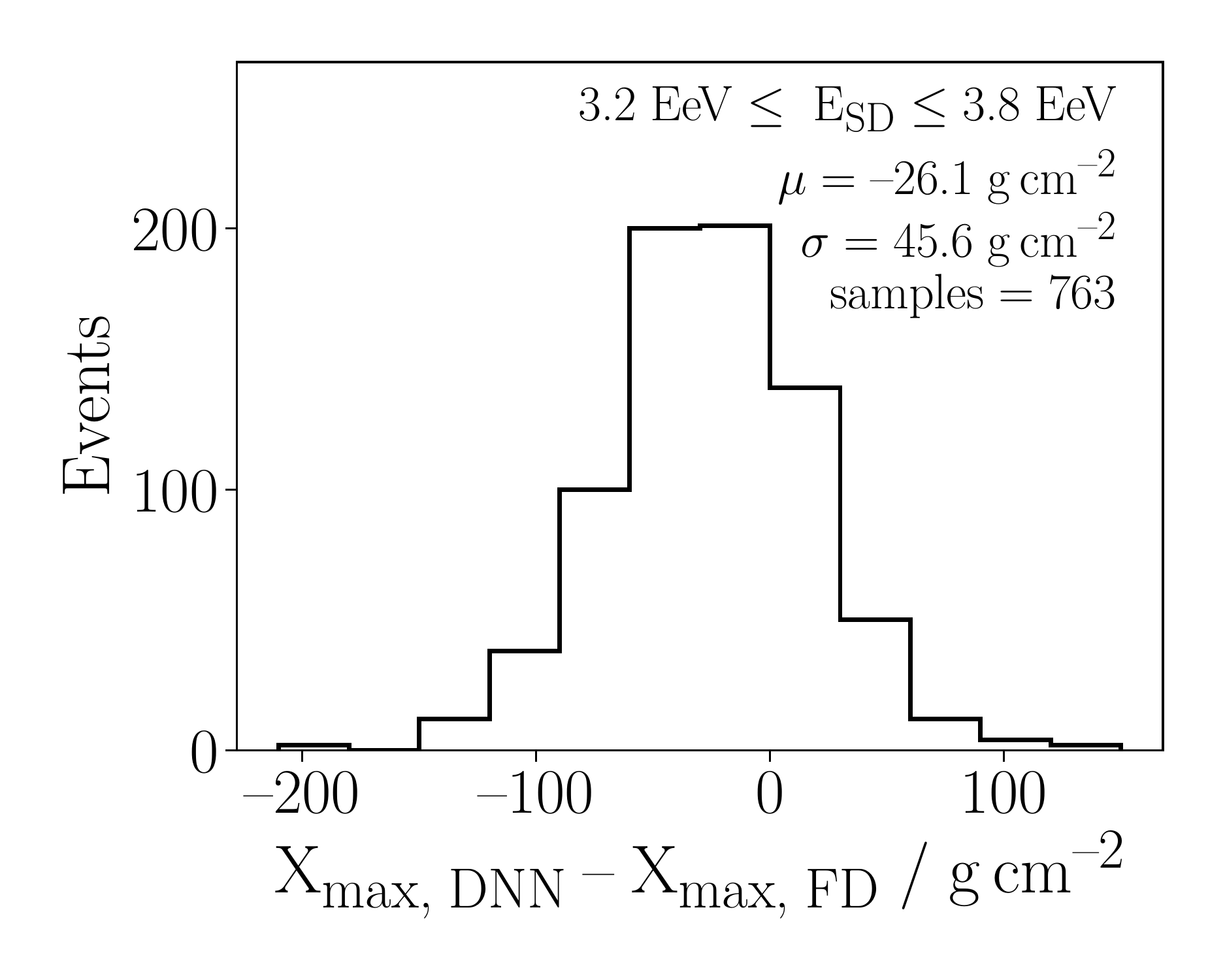}
                \label{fig:moments0}
            \end{centering}
        \end{subfigure}
        \begin{subfigure}[b]{0.325\textwidth}
            \begin{centering}
                \includegraphics[width=0.98\textwidth, trim={0.95cm 0 0.5cm 0}, clip]{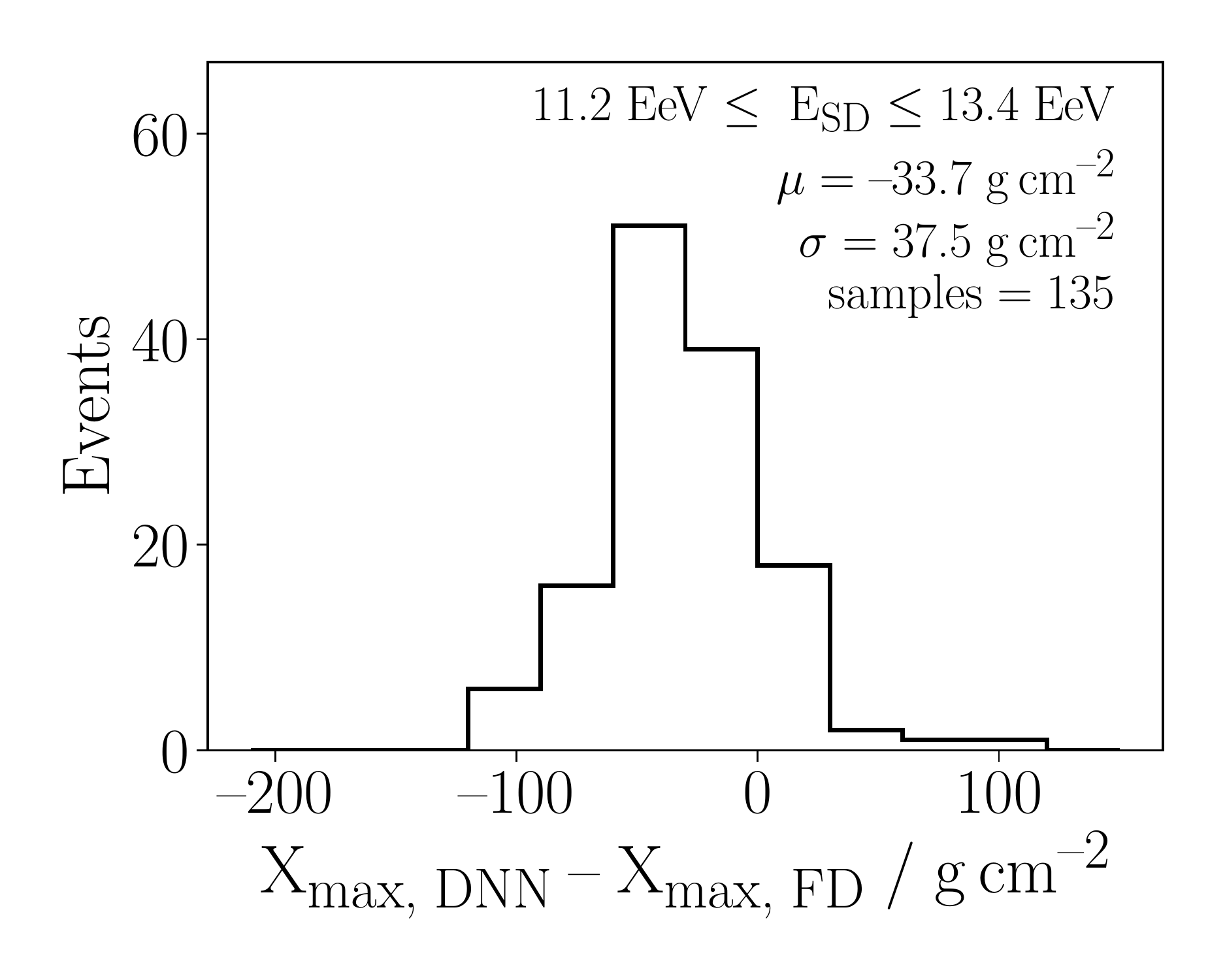}
                \label{fig:moments7}
            \end{centering}
        \end{subfigure}
        \begin{subfigure}[b]{0.325\textwidth}
            \begin{centering}
                \includegraphics[width=0.98\textwidth, trim={0.95cm 0 0.5cm 0}, clip]{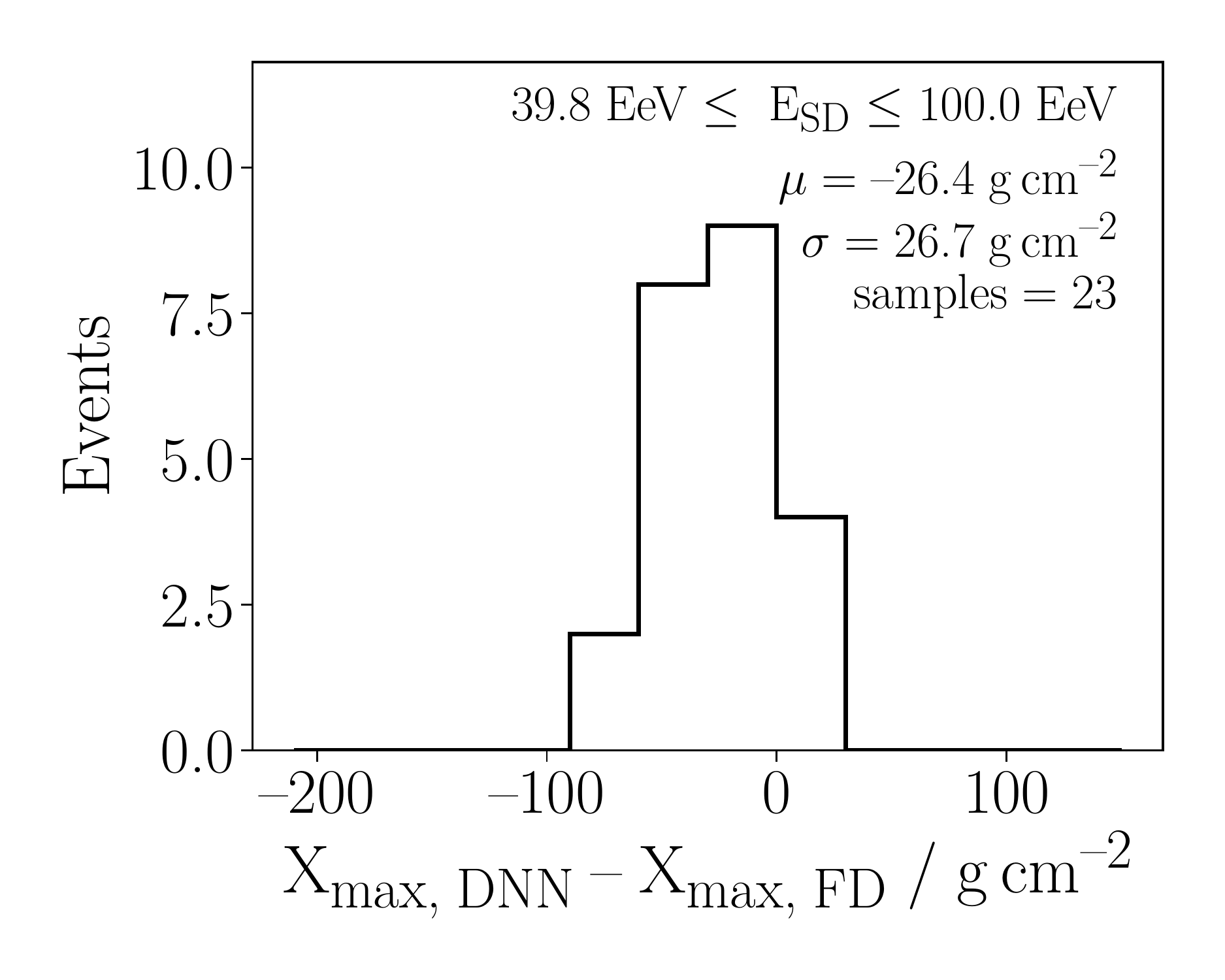}
                \label{fig:moments14}
            \end{centering}
        \end{subfigure}
    \caption{Difference of the \xmax values as reconstructed with the deep neural network and the fluorescence detector for three example SD-energy intervals.}
    \label{fig:reconstruction_distributions}
    \end{centering}
\end{figure*}

In Fig.~\ref{fig:reconstruction_distributions} we show the event-by-event difference $\Delta X_{\mathrm{max}}= X_{\mathrm{max,\;DNN}} - X_{\mathrm{max,\;FD}}$ for $3$ example energy intervals as reconstructed by the deep neural network and the FD.
All distributions are rather narrow and follow a Gaussian. This observation indicates that the shower-to-shower fluctuations measured with the FD, which feature a Gaussian convolved with an exponential~\cite{jose_xmax}, are also covered in the SD-based reconstruction.
The observed offset of around $-30$~\gcm indicates that the predicted \xmax values are too small on average.
Hence, the neural network predicts shallower shower maxima.

This shift is caused by slightly different shapes of the signal traces in simulations and data. One possible reason is the shower development. In fact, a bias was already present when evaluating the DNN using showers of hadronic interaction models different than used for the training. Also the size of the observed bias is not unexpected, since already previous analyses indicated problems of these models to precisely describe the lateral and longitudinal profiles of the muon component~\cite{Aab:2017cgk, auger_muons}.
In addition, differences in the detector simulation of the signal characteristics of the WCDs compared to data can contribute to the observed bias.

With modern methods, the differences between data and simulations can be quantitatively determined and eliminated~\cite{Erdmann:2018kuh}. In that data-driven method the simulation is refined to look more data-like using a semi-supervised learning approach.
In contrast, in this work we eliminate differences with a calibration using the independent measurements of the FD.

The energy dependency of this reconstruction bias is summarized in Fig.~\ref{fig:xmax_energy_data_bias}. Within the statistics of the hybrid data used, no significant energy dependency is observed. Only the lowest energy bin at $3$~EeV exhibits a slightly smaller bias.
Therefore, we calibrate the \xmax value reconstructed by the network with the average bias as measured with the FD:

\begin{align}
X_{\mathrm{max}}^{\mathrm{SD}} = X_{\mathrm{max,\;DNN}} + 30.0~\gcm.
\label{eq:calibration}
\end{align}
The offset was determined by a $1$-parameter fit to $30.0 \pm 0.6$~\gcm  and is shown by the horizontal fitted line in Fig.~\ref{fig:xmax_energy_data_bias}.
Even if a constant gives a rather good fit ($\chi^2 / \mathrm{ndf} = 1$), the first bins show an increased deviation from the FD measurements, which can be explained by biased reconstructions at low energies (compare Fig.~\ref{fig:zenith_energy_bias}).
If these bins are not taken into account, the fit remains stable and moves by roughly $-1$~\gcm.

\subsection{Potential for determining the second moment of the distribution of shower maxima}

In this section we will evaluate the resolution in the \xmax reconstruction of the deep neural network using the hybrid events.

\begin{figure*}[tb!]
    \begin{center}
        \begin{subfigure}[b]{0.495\textwidth} 
            \begin{centering}
                \includegraphics[width=0.99\textwidth]{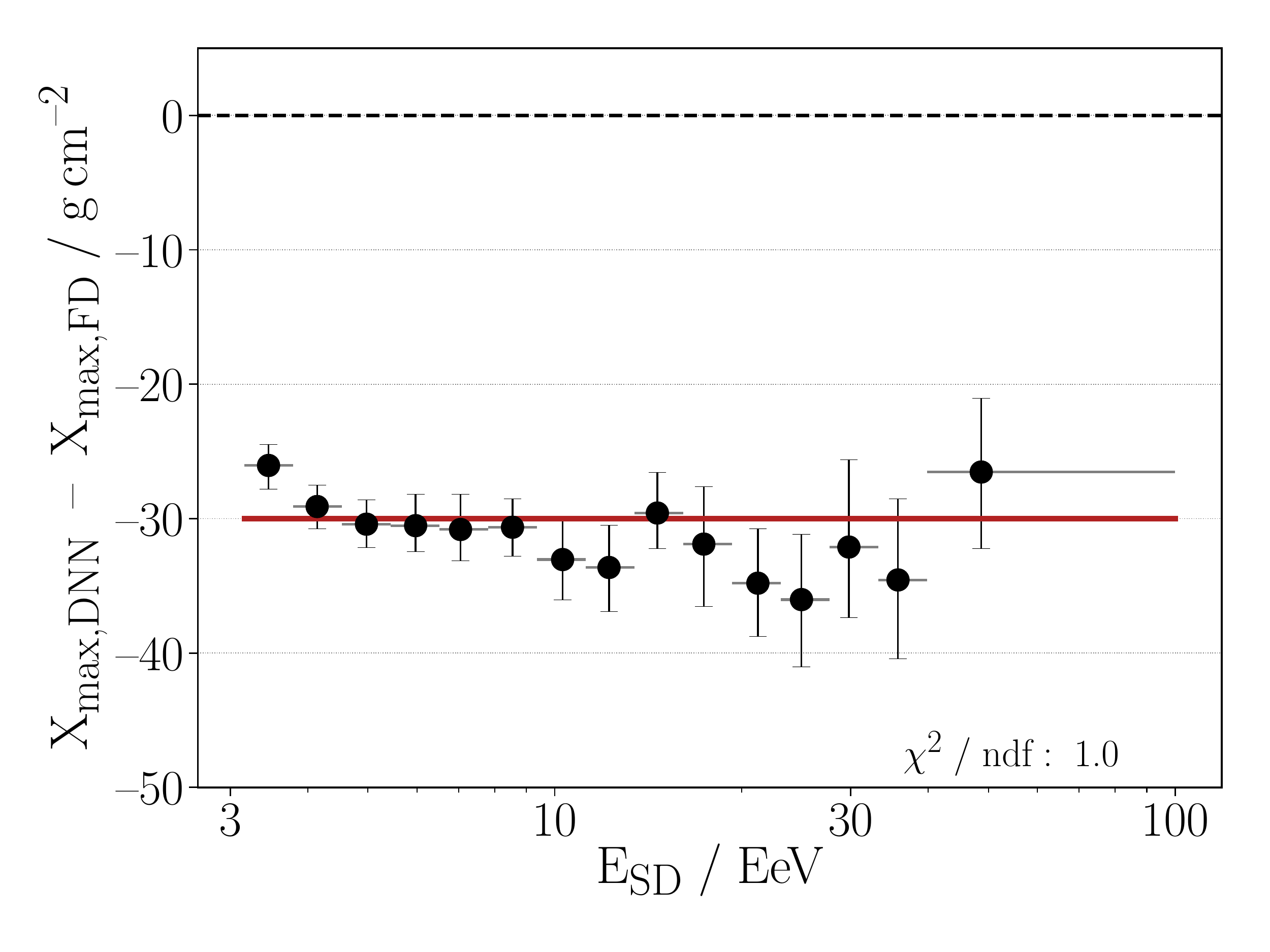}
                \subcaption{Bias}
                \label{fig:xmax_energy_data_bias}
            \end{centering}
        \end{subfigure}
        \begin{subfigure}[b]{0.495\textwidth} 
            \begin{centering}
                \includegraphics[width=0.99\textwidth]{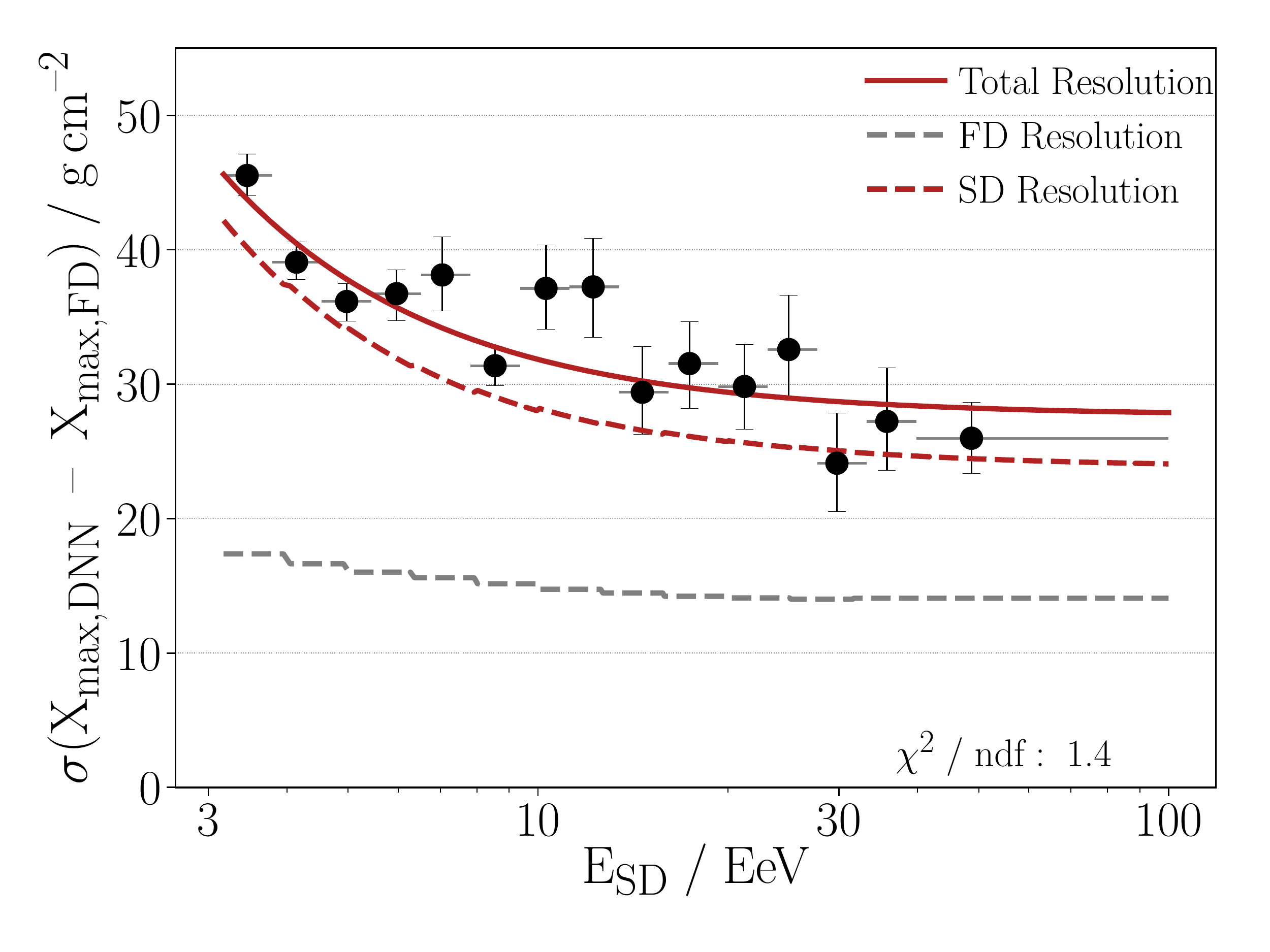}
                \subcaption{Resolution}
                \label{fig:xmax_energy_data_resolution}
            \end{centering}
        \end{subfigure}
        \caption{(a) Energy-dependent bias of the deep neural network with respect to the reconstruction of the fluorescence detector. (b) Energy-dependent resolution of the deep neural network with respect to the reconstruction of the fluorescence detector.}
        \label{fig:xmax_energy_data}
    \end{center}
\end{figure*}

In Fig.~\ref{fig:xmax_energy_data_resolution} we show the standard deviation $\sigma(\Delta X_{\mathrm{max}})$ of the event-wise differences between the \xmax value as reconstructed by the deep neural network and the \xmax value as measured by the FD. The symbols are located at the average energy of all events within the corresponding energy bin as indicated by the grey horizontal bars.

The vertical error bars indicate the uncertainty of $\sigma(\Delta X_{\mathrm{max}})$ combined for both reconstruction methods of the network and the FD. In detail, this uncertainty was calculated by a bootstrapping method using $1000$ random draws from the $\Delta X_{\mathrm{max}}$ distribution and calculating the standard deviation $\sigma$ for each set.

The energy dependency of the combined \xmax resolution is obtained by fitting an exponential function $\sigma_{\Delta X_{\mathrm{max}}}(E) = a\cdot e^{-b \cdot (\log_{10} E/\mathrm{eV}-18.5)} + c$ to the data, which is shown as a red curve.\newline
The values of the coefficients are $a=18.0\pm2.5$~\gcm, $b=2.9\pm1.2$ and $c=27.7\pm2.6$~\gcm. The \xmax resolution~\cite{fd_xmax} of the FD is indicated by the dashed grey curve. To obtain an estimate of the \xmax resolution of the deep neural network, we perform a quadratic subtraction of the FD resolution (dashed grey curve) from the combined \xmax resolution (solid red curve).
The resulting \xmax resolution of the deep neural network using only the measurements of the WCDs is shown as dashed red curve in Fig.~\ref{fig:xmax_energy_data_resolution}.
We find that the resolution on the measured data reaches less than $25$~\gcm above $20$~EeV, which is in the same order of magnitude as predicted by our simulation studies (Fig.~\ref{fig:EPOS_energy_sigma}) reinforcing the finding that the resolution seems to be independent of the hadronic model. 
This will enable new insights into the cosmic-ray composition at high energies.

\section{Summary}
In this work we presented a new approach for reconstructing the maximum shower depth \xmax using only the signal traces of the water-Cherenkov detectors (WCDs) placed on the ground, which record a tiny subset of the billions of shower particles.
It was shown that the presented method is capable of exploiting the data measured by the WCDs more comprehensively than ever before by adapting deep learning techniques, resulting in an unprecedented performance for mass composition studies using the surface detector (SD).
The new method yields both the large statistics of the SD and a measurement of \xmax on an event-by-event basis.
In consequence, our method opens up the possibility to measure the abundance of mass groups of UHECRs at $100$~EeV and beyond for the first time on data.

As reconstruction method, we have developed an advanced deep neural network which is especially suited for the situation of the Pierre Auger Observatory. The signal traces of the WCDs are analyzed by the network using so-called LSTM cells and their measurements are combined according to the hexagonal symmetry of the detector grid.

A key issue to correctly adjust the network parameters is the proper preparation of the data used for the network training. In addition to re-scaling and normalization of the signal amplitudes and time measurements, we implement real operation-conditions in the simulation data as data augmentation during the training. This includes missing WCDs due to hardware failures or showers falling close to the edges of the detector grid, missing signal traces of single photomultipliers and detector stations with saturated signal traces owing to high-energy events or very close shower cores. By including such effects, we make the network robust against small differences between simulation and measured data, enhancing its generalization capacities and providing an accurate reconstruction of \xmax for zenith angles up to $60^{\circ}$ and even for events with saturated station electronics.

Initially we evaluate the performance of the network on simulated data.
When evaluating the network using disjunct data from the same simulation as used for training, we observe an almost bias free reconstruction of \xmax. The \xmax resolution improves with increasing cosmic ray energy and is composition dependent. For proton-induced showers the resolution is $38$~\gcm at $10$~EeV and improves to $28$~\gcm at the highest energies. For iron primaries the resolution is better than $20$~\gcm above $20$~EeV.

When changing the hadronic interaction model for the evaluation of the network, we find that only the absolute bias in the \xmax reconstruction changes. In contrast to this shift in \xmax, the resolution of \xmax appears to be essentially independent of the hadronic interaction model. Additionally, we found that the network is able to reproduce the correct shape of the \xmax distribution for all elements and hadronic interaction models at energies above $10$~EeV.

Finally, we test the network performance using events which include measurements of the fluorescence detector. We first eliminated effects of detector aging from long-term operation of the observatory. Compared to the results of the fluorescence measurement a remaining shift of the reconstructed \xmax value of the network of about $30$~\gcm was found. The shift is independent of the cosmic ray energy allowing for a straightforward calibration.
The observed shift is larger than the constant shift of up to $-15$~\gcm observed in the application of different interaction models. However, since all these models are not able to precisely describe the muonic component of air showers, a shift in the observed magnitude is not unexpected. Additionally, inaccuracies in the detector simulation could contribute to the observed bias.

We then estimate the \xmax resolution of the network on data by subtracting the \xmax resolution of the fluorescence detector in quadrature. We obtain a resolution of about $25$~\gcm above $20$~EeV which is well compatible with the resolution expected from our simulation studies.

Thus, a high statistics measurement of the first moment $\langle X_{\mathrm{max}}\rangle$ and the second moment $\sigma(X_{\mathrm{max}})$ using the deep neural network reconstruction of the WCD-signal traces has a great potential to provide new insights into the cosmic-ray composition at the highest energies.

\newpage

\section*{Acknowledgments}

\begin{sloppypar}
The successful installation, commissioning, and operation of the Pierre
Auger Observatory would not have been possible without the strong
commitment and effort from the technical and administrative staff in
Malarg\"ue. We are very grateful to the following agencies and
organizations for financial support:
\end{sloppypar}

\begin{sloppypar}
Argentina -- Comisi\'on Nacional de Energ\'\i{}a At\'omica; Agencia Nacional de
Promoci\'on Cient\'\i{}fica y Tecnol\'ogica (ANPCyT); Consejo Nacional de
Investigaciones Cient\'\i{}ficas y T\'ecnicas (CONICET); Gobierno de la
Provincia de Mendoza; Municipalidad de Malarg\"ue; NDM Holdings and Valle
Las Le\~nas; in gratitude for their continuing cooperation over land
access; Australia -- the Australian Research Council; Brazil -- Conselho
Nacional de Desenvolvimento Cient\'\i{}fico e Tecnol\'ogico (CNPq);
Financiadora de Estudos e Projetos (FINEP); Funda\c{c}\~ao de Amparo \`a
Pesquisa do Estado de Rio de Janeiro (FAPERJ); S\~ao Paulo Research
Foundation (FAPESP) Grants No.~2019/10151-2, No.~2010/07359-6 and
No.~1999/05404-3; Minist\'erio da Ci\^encia, Tecnologia, Inova\c{c}\~oes e
Comunica\c{c}\~oes (MCTIC); Czech Republic -- Grant No.~MSMT CR LTT18004,
LM2015038, LM2018102, CZ.02.1.01/0.0/0.0/16{\textunderscore}013/0001402,
CZ.02.1.01/0.0/0.0/18{\textunderscore}046/0016010 and
CZ.02.1.01/0.0/0.0/17{\textunderscore}049/0008422; France -- Centre de Calcul
IN2P3/CNRS; Centre National de la Recherche Scientifique (CNRS); Conseil
R\'egional Ile-de-France; D\'epartement Physique Nucl\'eaire et Corpusculaire
(PNC-IN2P3/CNRS); D\'epartement Sciences de l'Univers (SDU-INSU/CNRS);
Institut Lagrange de Paris (ILP) Grant No.~LABEX ANR-10-LABX-63 within
the Investissements d'Avenir Programme Grant No.~ANR-11-IDEX-0004-02;
Germany -- Bundesministerium f\"ur Bildung und Forschung (BMBF); Deutsche
Forschungsgemeinschaft (DFG); Finanzministerium Baden-W\"urttemberg;
Helmholtz Alliance for Astroparticle Physics (HAP);
Helmholtz-Gemeinschaft Deutscher Forschungszentren (HGF); Ministerium
f\"ur Innovation, Wissenschaft und Forschung des Landes
Nordrhein-Westfalen; Ministerium f\"ur Wissenschaft, Forschung und Kunst
des Landes Baden-W\"urttemberg; Italy -- Istituto Nazionale di Fisica
Nucleare (INFN); Istituto Nazionale di Astrofisica (INAF); Ministero
dell'Istruzione, dell'Universit\'a e della Ricerca (MIUR); CETEMPS Center
of Excellence; Ministero degli Affari Esteri (MAE); M\'exico -- Consejo
Nacional de Ciencia y Tecnolog\'\i{}a (CONACYT) No.~167733; Universidad
Nacional Aut\'onoma de M\'exico (UNAM); PAPIIT DGAPA-UNAM; The Netherlands
-- Ministry of Education, Culture and Science; Netherlands Organisation
for Scientific Research (NWO); Dutch national e-infrastructure with the
support of SURF Cooperative; Poland -Ministry of Science and Higher
Education, grant No.~DIR/WK/2018/11; National Science Centre, Grants
No.~2013/08/M/ST9/00322, No.~2016/23/B/ST9/01635 and No.~HARMONIA
5--2013/10/M/ST9/00062, UMO-2016/22/M/ST9/00198; Portugal -- Portuguese
national funds and FEDER funds within Programa Operacional Factores de
Competitividade through Funda\c{c}\~ao para a Ci\^encia e a Tecnologia
(COMPETE); Romania -- Romanian Ministry of Education and Research, the
Program Nucleu within MCI (PN19150201/16N/2019 and PN19060102) and
project PN-III-P1-1.2-PCCDI-2017-0839/19PCCDI/2018 within PNCDI III;
Slovenia -- Slovenian Research Agency, grants P1-0031, P1-0385, I0-0033,
N1-0111; Spain -- Ministerio de Econom\'\i{}a, Industria y Competitividad
(FPA2017-85114-P and PID2019-104676GB-C32, Xunta de Galicia (ED431C
2017/07), Junta de Andaluc\'\i{}a (SOMM17/6104/UGR, P18-FR-4314) Feder Funds,
RENATA Red Nacional Tem\'atica de Astropart\'\i{}culas (FPA2015-68783-REDT) and
Mar\'\i{}a de Maeztu Unit of Excellence (MDM-2016-0692); USA -- Department of
Energy, Contracts No.~DE-AC02-07CH11359, No.~DE-FR02-04ER41300,
No.~DE-FG02-99ER41107 and No.~DE-SC0011689; National Science Foundation,
Grant No.~0450696; The Grainger Foundation; Marie Curie-IRSES/EPLANET;
European Particle Physics Latin American Network; and UNESCO.
\end{sloppypar}

\begin{center}
\rule{0.1\columnwidth}{0.5pt}\,\raisebox{-0.5pt}{\rule{0.05\columnwidth}{1.5pt}}~\raisebox{-0.375ex}{\scriptsize$\bullet$}~\raisebox{-0.5pt}{\rule{0.05\columnwidth}{1.5pt}}\,\rule{0.1\columnwidth}{0.5pt}
\end{center}

\section*{The Pierre Auger Collaboration}

A.~Aab$^{80}$,
P.~Abreu$^{72}$,
M.~Aglietta$^{52,50}$,
J.M.~Albury$^{12}$,
I.~Allekotte$^{1}$,
A.~Almela$^{8,11}$,
J.~Alvarez-Mu\~niz$^{79}$,
R.~Alves Batista$^{80}$,
G.A.~Anastasi$^{61,50}$,
L.~Anchordoqui$^{87}$
% ...

{\footnotesize

\begin{description}[labelsep=0.2em,align=right,labelwidth=0.7em,labelindent=0em,leftmargin=2em,noitemsep]
\item[$^{1}$] Centro At\'omico Bariloche and Instituto Balseiro (CNEA-UNCuyo-CONICET), San Carlos de Bariloche, Argentina
\item[$^{2}$] Centro de Investigaciones en L\'aseres y Aplicaciones, CITEDEF and CONICET, Villa Martelli, Argentina
\item[$^{3}$] Departamento de F\'\i{}sica and Departamento de Ciencias de la Atm\'osfera y los Oc\'eanos, FCEyN, Universidad de Buenos Aires and CONICET, Buenos Aires, Argentina
\item[$^{4}$] IFLP, Universidad Nacional de La Plata and CONICET, La Plata, Argentina
\item[$^{5}$] Instituto de Astronom\'\i{}a y F\'\i{}sica del Espacio (IAFE, CONICET-UBA), Buenos Aires, Argentina
\item[$^{6}$] Instituto de F\'\i{}sica de Rosario (IFIR) -- CONICET/U.N.R.\ and Facultad de Ciencias Bioqu\'\i{}micas y Farmac\'euticas U.N.R., Rosario, Argentina
\item[$^{7}$] Instituto de Tecnolog\'\i{}as en Detecci\'on y Astropart\'\i{}culas (CNEA, CONICET, UNSAM), and Universidad Tecnol\'ogica Nacional -- Facultad Regional Mendoza (CONICET/CNEA), Mendoza, Argentina
\item[$^{8}$] Instituto de Tecnolog\'\i{}as en Detecci\'on y Astropart\'\i{}culas (CNEA, CONICET, UNSAM), Buenos Aires, Argentina
\item[$^{9}$] Observatorio Pierre Auger, Malarg\"ue, Argentina
\item[$^{10}$] Observatorio Pierre Auger and Comisi\'on Nacional de Energ\'\i{}a At\'omica, Malarg\"ue, Argentina
\item[$^{11}$] Universidad Tecnol\'ogica Nacional -- Facultad Regional Buenos Aires, Buenos Aires, Argentina
\item[$^{12}$] University of Adelaide, Adelaide, S.A., Australia
\item[$^{13}$] Universit\'e Libre de Bruxelles (ULB), Brussels, Belgium
\item[$^{14}$] Vrije Universiteit Brussels, Brussels, Belgium
\item[$^{15}$] Centro Brasileiro de Pesquisas Fisicas, Rio de Janeiro, RJ, Brazil
\item[$^{16}$] Centro Federal de Educa\c{c}\~ao Tecnol\'ogica Celso Suckow da Fonseca, Nova Friburgo, Brazil
\item[$^{17}$] Universidade de S\~ao Paulo, Escola de Engenharia de Lorena, Lorena, SP, Brazil
\item[$^{18}$] Universidade de S\~ao Paulo, Instituto de F\'\i{}sica de S\~ao Carlos, S\~ao Carlos, SP, Brazil
\item[$^{19}$] Universidade de S\~ao Paulo, Instituto de F\'\i{}sica, S\~ao Paulo, SP, Brazil
\item[$^{20}$] Universidade Estadual de Campinas, IFGW, Campinas, SP, Brazil
\item[$^{21}$] Universidade Estadual de Feira de Santana, Feira de Santana, Brazil
\item[$^{22}$] Universidade Federal do ABC, Santo Andr\'e, SP, Brazil
\item[$^{23}$] Universidade Federal do Paran\'a, Setor Palotina, Palotina, Brazil
\item[$^{24}$] Universidade Federal do Rio de Janeiro, Instituto de F\'\i{}sica, Rio de Janeiro, RJ, Brazil
\item[$^{25}$] Universidade Federal do Rio de Janeiro (UFRJ), Observat\'orio do Valongo, Rio de Janeiro, RJ, Brazil
\item[$^{26}$] Universidade Federal Fluminense, EEIMVR, Volta Redonda, RJ, Brazil
\item[$^{27}$] Universidad de Medell\'\i{}n, Medell\'\i{}n, Colombia
\item[$^{28}$] Universidad Industrial de Santander, Bucaramanga, Colombia
\item[$^{29}$] Charles University, Faculty of Mathematics and Physics, Institute of Particle and Nuclear Physics, Prague, Czech Republic
\item[$^{30}$] Institute of Physics of the Czech Academy of Sciences, Prague, Czech Republic
\item[$^{31}$] Palacky University, RCPTM, Olomouc, Czech Republic
\item[$^{32}$] CNRS/IN2P3, IJCLab, Universit\'e Paris-Saclay, Orsay, France
\item[$^{33}$] Laboratoire de Physique Nucl\'eaire et de Hautes Energies (LPNHE), Sorbonne Universit\'e, Universit\'e de Paris, CNRS-IN2P3, Paris, France
\item[$^{34}$] Univ.\ Grenoble Alpes, CNRS, Grenoble Institute of Engineering Univ.\ Grenoble Alpes, LPSC-IN2P3, 38000 Grenoble, France
\item[$^{35}$] Universit\'e Paris-Saclay, CNRS/IN2P3, IJCLab, Orsay, France
\item[$^{36}$] Bergische Universit\"at Wuppertal, Department of Physics, Wuppertal, Germany
\item[$^{37}$] Karlsruhe Institute of Technology (KIT), Institute for Experimental Particle Physics, Karlsruhe, Germany
\item[$^{38}$] Karlsruhe Institute of Technology (KIT), Institut f\"ur Prozessdatenverarbeitung und Elektronik, Karlsruhe, Germany
\item[$^{39}$] Karlsruhe Institute of Technology (KIT), Institute for Astroparticle Physics, Karlsruhe, Germany
\item[$^{40}$] RWTH Aachen University, III.\ Physikalisches Institut A, Aachen, Germany
\item[$^{41}$] Universit\"at Hamburg, II.\ Institut f\"ur Theoretische Physik, Hamburg, Germany
\item[$^{42}$] Universit\"at Siegen, Department Physik -- Experimentelle Teilchenphysik, Siegen, Germany
\item[$^{43}$] Gran Sasso Science Institute, L'Aquila, Italy
\item[$^{44}$] INFN Laboratori Nazionali del Gran Sasso, Assergi (L'Aquila), Italy
\item[$^{45}$] INFN, Sezione di Catania, Catania, Italy
\item[$^{46}$] INFN, Sezione di Lecce, Lecce, Italy
\item[$^{47}$] INFN, Sezione di Milano, Milano, Italy
\item[$^{48}$] INFN, Sezione di Napoli, Napoli, Italy
\item[$^{49}$] INFN, Sezione di Roma ``Tor Vergata'', Roma, Italy
\item[$^{50}$] INFN, Sezione di Torino, Torino, Italy
\item[$^{51}$] Istituto di Astrofisica Spaziale e Fisica Cosmica di Palermo (INAF), Palermo, Italy
\item[$^{52}$] Osservatorio Astrofisico di Torino (INAF), Torino, Italy
\item[$^{53}$] Politecnico di Milano, Dipartimento di Scienze e Tecnologie Aerospaziali , Milano, Italy
\item[$^{54}$] Universit\`a del Salento, Dipartimento di Matematica e Fisica ``E.\ De Giorgi'', Lecce, Italy
\item[$^{55}$] Universit\`a dell'Aquila, Dipartimento di Scienze Fisiche e Chimiche, L'Aquila, Italy
\item[$^{56}$] Universit\`a di Catania, Dipartimento di Fisica e Astronomia, Catania, Italy
\item[$^{57}$] Universit\`a di Milano, Dipartimento di Fisica, Milano, Italy
\item[$^{58}$] Universit\`a di Napoli ``Federico II'', Dipartimento di Fisica ``Ettore Pancini'', Napoli, Italy
\item[$^{59}$] Universit\`a di Palermo, Dipartimento di Fisica e Chimica ''E.\ Segr\`e'', Palermo, Italy
\item[$^{60}$] Universit\`a di Roma ``Tor Vergata'', Dipartimento di Fisica, Roma, Italy
\item[$^{61}$] Universit\`a Torino, Dipartimento di Fisica, Torino, Italy
\item[$^{62}$] Benem\'erita Universidad Aut\'onoma de Puebla, Puebla, M\'exico
\item[$^{63}$] Centro de Investigaci\'on y de Estudios Avanzados del IPN (CINVESTAV), M\'exico, D.F., M\'exico
\item[$^{64}$] Unidad Profesional Interdisciplinaria en Ingenier\'\i{}a y Tecnolog\'\i{}as Avanzadas del Instituto Polit\'ecnico Nacional (UPIITA-IPN), M\'exico, D.F., M\'exico
\item[$^{65}$] Universidad Aut\'onoma de Chiapas, Tuxtla Guti\'errez, Chiapas, M\'exico
\item[$^{66}$] Universidad Michoacana de San Nicol\'as de Hidalgo, Morelia, Michoac\'an, M\'exico
\item[$^{67}$] Universidad Nacional Aut\'onoma de M\'exico, M\'exico, D.F., M\'exico
\item[$^{68}$] Universidad Nacional de San Agustin de Arequipa, Facultad de Ciencias Naturales y Formales, Arequipa, Peru
\item[$^{69}$] Institute of Nuclear Physics PAN, Krakow, Poland
\item[$^{70}$] University of \L{}\'od\'z, Faculty of Astrophysics, \L{}\'od\'z, Poland
\item[$^{71}$] University of \L{}\'od\'z, Faculty of High-Energy Astrophysics,\L{}\'od\'z, Poland
\item[$^{72}$] Laborat\'orio de Instrumenta\c{c}\~ao e F\'\i{}sica Experimental de Part\'\i{}culas -- LIP and Instituto Superior T\'ecnico -- IST, Universidade de Lisboa -- UL, Lisboa, Portugal
\item[$^{73}$] ``Horia Hulubei'' National Institute for Physics and Nuclear Engineering, Bucharest-Magurele, Romania
\item[$^{74}$] Institute of Space Science, Bucharest-Magurele, Romania
\item[$^{75}$] University Politehnica of Bucharest, Bucharest, Romania
\item[$^{76}$] Center for Astrophysics and Cosmology (CAC), University of Nova Gorica, Nova Gorica, Slovenia
\item[$^{77}$] Experimental Particle Physics Department, J.\ Stefan Institute, Ljubljana, Slovenia
\item[$^{78}$] Universidad de Granada and C.A.F.P.E., Granada, Spain
\item[$^{79}$] Instituto Galego de F\'\i{}sica de Altas Enerx\'\i{}as (IGFAE), Universidade de Santiago de Compostela, Santiago de Compostela, Spain
\item[$^{80}$] IMAPP, Radboud University Nijmegen, Nijmegen, The Netherlands
\item[$^{81}$] KVI -- Center for Advanced Radiation Technology, University of Groningen, Groningen, The Netherlands
\item[$^{82}$] Nationaal Instituut voor Kernfysica en Hoge Energie Fysica (NIKHEF), Science Park, Amsterdam, The Netherlands
\item[$^{83}$] Stichting Astronomisch Onderzoek in Nederland (ASTRON), Dwingeloo, The Netherlands
\item[$^{84}$] Universiteit van Amsterdam, Faculty of Science, Amsterdam, The Netherlands
\item[$^{85}$] Case Western Reserve University, Cleveland, OH, USA
\item[$^{86}$] Colorado School of Mines, Golden, CO, USA
\item[$^{87}$] Department of Physics and Astronomy, Lehman College, City University of New York, Bronx, NY, USA
\item[$^{88}$] Louisiana State University, Baton Rouge, LA, USA
\item[$^{89}$] Michigan Technological University, Houghton, MI, USA
\item[$^{90}$] New York University, New York, NY, USA
\item[$^{91}$] Pennsylvania State University, University Park, PA, USA
\item[$^{92}$] University of Chicago, Enrico Fermi Institute, Chicago, IL, USA
\item[$^{93}$] University of Delaware, Department of Physics and Astronomy, Bartol Research Institute, Newark, DE, USA
\item[$^{94}$] University of Wisconsin-Madison, Department of Physics and WIPAC, Madison, WI, USA
\item[] -----
\item[$^{a}$] Fermi National Accelerator Laboratory, Fermilab, Batavia, IL, USA
\item[$^{b}$] Max-Planck-Institut f\"ur Radioastronomie, Bonn, Germany
\item[$^{c}$] School of Physics and Astronomy, University of Leeds, Leeds, United Kingdom
\item[$^{d}$] Colorado State University, Fort Collins, CO, USA
\item[$^{e}$] now at Hakubi Center for Advanced Research and Graduate School of Science, Kyoto University, Kyoto, Japan
\item[$^{f}$] also at University of Bucharest, Physics Department, Bucharest, Romania
\end{description}

}

\end{document}